\documentclass[]{aastex631}

\usepackage{amsmath}
\usepackage{color}
\usepackage{float}
\usepackage{bm}
\usepackage{soul}
\usepackage{xcolor}
\usepackage{orcidlink}

\shorttitle{Linear mode conversion theory}
\shortauthors{Krafft and Volokitin}

\begin{document}

\title{Linear mode conversion theory of radio emission from turbulent solar wind plasmas}

\correspondingauthor{Catherine Krafft}
\email{catherine.krafft@universite-paris-saclay.fr}

\author{C. Krafft\orcidlink{0000-0002-8595-4772}}
\affiliation{Laboratoire de Physique des Plasmas (LPP), CNRS, Sorbonne Université, \\ 
Observatoire de Paris, Université Paris-Saclay,\\ Ecole polytechnique, Institut Polytechnique de Paris, 91128 Palaiseau, France}
\affiliation{Institut Universitaire de France (IUF)}

\author{A. S. Volokitin\orcidlink{0009-0005-3143-6282}}
\affiliation{Laboratoire de Physique des Plasmas (LPP), CNRS, Sorbonne Université, \\ 
Observatoire de Paris, Université Paris-Saclay,\\ Ecole polytechnique, Institut Polytechnique de Paris, 91128 Palaiseau, France}
\affiliation{Pushkov Institute of Terrestrial Magnetism, Ionosphere and Radio Wave Propagation, Troitsk, 142190 Moscow, Russia}

\begin{abstract}
 
This work presents a new theoretical and numerical model describing all possible linear interactions between upper-hybrid wave turbulence and random density fluctuations in a solar wind plasma; not only linear processes as wave reflection, refraction, scattering, tunneling, trapping, or mode conversion at constant frequency are taken into account, but also linear wave coupling, interferences between scattered waves, etc. Compact equations describing the time evolution of electromagnetic fields radiated in the $\mathcal{O}$, $\mathcal{X}$ and $\mathcal{Z}$ modes by the current  due to transformations of upper-hybrid waves on density fluctuations, as well as the dispersion and polarization properties of the modes, are determined analytically and solved numerically, providing the time variations of electromagnetic energies and corresponding radiation rates. Jointly, on the basis of these numerical results that validate theoretical hypotheses, analytical calculations are conducted in the framework of weak turbulence theory extended to randomly inhomogeneous plasmas, that recover the main physical conclusions stated using the new model. The dependencies of radiation rates on plasma parameters as the magnetization, the electron thermal velocity and the average level of random density fluctuations are determined in the form of scaling laws. This work opens a new way to analyze the efficiency of electromagnetic emissions at plasma frequency by realistic wave and density turbulence spectra interacting in solar wind plasmas. 
\end{abstract}

	\section{Introduction}
Type III solar radio bursts have been routinely observed since decades in the interplanetary space by spacecraft and ground-based radiotelescopes (e.g. \cite{Dulk1985}, \cite{ReidRatcliffe2014}, and references therein). Energetic electron beams ejected during solar flares and propagating along open magnetic field lines generate upper-hybrid wave turbulence that in turn radiates electromagnetic waves at the electron plasma frequency $\omega_{p}$ and its harmonics, via successive linear and nonlinear processes. Today, satellites such as Parker Solar Probe (\cite{Fox2016}) and Solar Orbiter (\cite{Muller2020}), as well as radiotelescopes as the Low Frequency Array/LOFAR (\cite{VanHaarlem2013})  provide a large amount of new observations on electromagnetic wave emission and beam radiation during type III solar radio bursts (e.g. \cite{Chen2021},  \cite{ThejappaMacDowall2021}, \cite{ReidKontar2021}, \cite{Badman2022}, \cite{Jebaraj2023a}, \cite{Lorfing2023}, \cite{Krupar2024a}, \cite{Krupar2024b}).
    
    Whereas electromagnetic wave emissions at $\omega_{p}$ were first considered to result from the scattering of Langmuir waves off thermal ions (\cite{GinzburgZheleznyakov1958}), different approaches were further proposed to explain their generation mechanisms. Some authors suggested, in the framework of weak turbulence theory, that nonlinear wave decay or coalescence processes $\mathcal{L}  \rightarrow \mathcal{O}\pm \mathcal{S}$ involving Langmuir waves $\mathcal{L}$ and ion acoustic waves $\mathcal{S}$ can be responsible for radiation at $\omega_{p}$ of electromagnetic ordinary waves $\mathcal{O}$ (e.g. \cite{Tsytovich1970}, \cite{Melrose1980}). Other works invoked the theory of strong turbulence (\cite{Papadopoulos1974}), or proposed the antenna mechanism where electromagnetic emissions can be radiated by density cavities containing trapped Langmuir waves (\cite{Malaspina2012}). As random density fluctuations $\delta n$ of average levels $\Delta N=\langle (\delta n/n_0)^2\rangle^{1/2}$ of a few percent of the plasma density $n_0$, which were observed in the solar wind (e.g. \cite{ Celnikier1983},  \cite{Krupar2015}, \cite{Krupar2020}), interact with Langmuir wave turbulence generated by electron beams (\cite{NishikawaRyutov1976}, \cite{Muschietti1985}), transformation processes of electrostatic waves on plasma inhomogeneities and, in particular, their linear mode conversion (LMC) at constant frequency, were proposed to explain electromagnetic radiation at $\omega_{p}$. Such processes were studied analytically and numerically considering monochromatic waves incident on density gradients (\cite{Hinkel-Lipsker1989}, \cite{Hinkel-Lipsker1991}, \cite{CairnsWilles2005}) or wave turbulence scattering on external random density fluctuations (\cite{VolokitinKrafft2018}, \cite{Krasnoselskikh2019}, \cite{VolokitinKrafft2020}, \cite{KrafftSavoini2022a},  \cite{KrafftSavoini2024}, \cite{Krafft2024}, \cite{Krafft2025}).
    
        Furthermore, electromagnetic emissions at $\omega_{p}$ were studied analytically and numerically using different approaches and modeling. Some authors solved the weak turbulence or quasi-linear equations in homogeneous plasmas and calculated wave emission due to three-wave decay (e.g. \cite{EdneyRobinson1999}, \cite{Li2005}, \cite{Ziebell2015}, \cite{Lee2019}). On this basis, models were built to describe the injection of an electron beam in a plasma source with small-scale density fluctuations, the radiation of electromagnetic waves at $\omega_{p}$, their escape away from their generation region, and their propagation along a decreasing plasma density profile (\cite{Li2008a}, \cite{Li2008b}, \cite{RatcliffeReid2014}). Other approaches, involving randomly inhomogeneous plasmas, were developed to study electromagnetic radiation resulting from Langmuir wave transformations on density fluctuations, i.e. their reflection on density inhomogeneities resulting in their partial conversion into electromagnetic energy (\cite{Krasnoselskikh2019}), or the determination of electromagnetic radiation rates using Zakharov equations coupled with a modified theory of retarded potentials (\cite{VolokitinKrafft2018}, \cite{VolokitinKrafft2020}). Finally, electromagnetic radiation at $\omega_{p}$ was studied more recently within the framework of two-dimensional (2D) Particle-In-Cell (PIC) simulations (e.g. \cite{Rhee2009}, \cite{Lee2022}, \cite{KrafftSavoini2022a}, \cite{Krafft2024}, \cite{Polanco2025a}, \cite{Krafft2025}, and references therein).
    
    With recent solar missions such as Parker Solar Probe and Solar Orbiter, that are now approaching closer to the Sun, theoretical and numerical studies involving magnetized plasmas are becoming essential. However, almost all work on plasma emission and related issues has been to date carried out in the approximation of unmagnetized plasmas. Note, however, some (not exhaustive) examples of theoretical and numerical studies considering magnetized plasmas, performed on nonlinear Langmuir wave decay (\cite{Akimoto1989}, \cite{Layden2013}, \cite{CairnsLayden2018}), linear mode conversion (\cite{Yin1998}, \cite{Kim2007}, \cite{Kim2008}, \cite{Schleyer2013}, \cite{Schleyer2014}, \cite{Krafft2025}), electromagnetic wave polarization (\cite{MelroseSy1972}, \cite{Zlotnik1981}, \cite{WillesCairns2000}), or using Particle-In-Cell (PIC) simulations (\cite{DumNishikawa1994}, \cite{Zhou2020}, \cite{Lee2022}, \cite{Polanco2025a}). In this regard, the present work considers randomly inhomogeneous and weakly magnetized plasmas (with $\omega_{c}/\omega_{p}\lesssim 0.2$, where $\omega_{c}$ is the electron cyclotron frequency). More specifically we study, in such a plasma, the evolution of upper-hybrid wave turbulence  and its electromagnetic radiation at $\omega_{p}$. The main objective is to demonstrate the essential impact of plasma density inhomogeneities and magnetization on the radiation rates and the spectral distributions of electromagnetic emissions in the ordinary $\mathcal{O}$-mode, as well as in the fast and slow extraordinary modes $\mathcal{X}$ and $\mathcal{Z}$, due to upper-hybrid wave transformations on density irregularities as, for example, linear mode conversion at frequency $\omega_p$.
    
    One of the mechanisms that generates electromagnetic waves at $\omega_{p}$ in a turbulent plasma is the interaction between high- and low-frequency oscillations or, almost the same, the scattering of high-frequency waves on density fluctuations. As usually thought, especially in the framework of weak turbulence theory, density fluctuations required in scattering processes arise from nonlinear processes of wave decay and coalescence. However, attention has recently been  drawn to the fact that, according to observations (e.g. \cite{Celnikier1983}, \cite{Krupar2015}, \cite{Krupar2020}), density fluctuations can exist in the solar wind independently of nonlinear processes involving high-frequency waves, and that their amplitudes can significantly exceed the levels expected by the weak turbulence theory. That is the context assumed in this paper. Based on a new theoretical and numerical model involving two-dimensional Zakharov equations in a weakly magnetized plasma, this work provides compact equations governing wave radiation emitted at $\omega_{p}$ in ordinary and extraordinary electromagnetic modes by turbulent upper-hybrid waves. Radiation rates in each mode are determined, as well as their dependence on the average level of density fluctuations $\Delta N$, the ratio $\omega_{c}/\omega_{p}$ of the cyclotron to the plasma frequency, and the ratio $v_{T}/c$ of the electron thermal to  the light  velocity.
    
    Note that if the plasma radio source is optically thick, the problem becomes considerably more complicated as it is necessary, in order to calculate the energy flux carried by electromagnetic waves escaping from the source, to take into account re-emission and absorption processes. But in many cases of practical  interest, such as in the solar wind, the radiating source is optically thin, and we can limit ourselves to calculating the local rate of electromagnetic wave radiation in a given volume, assuming that all these waves freely leave their source and propagate further away. In this case, it is sufficient to determine the rate of transformation of electrostatic wave energy into electromagnetic energy.

\section{Theoretical and numerical model}

\subsection{Description of the radio source}
 In order to calculate electromagnetic wave radiation by a plasma source, we generalize an
approach developed in our previous works (\cite{VolokitinKrafft2018}, \cite{VolokitinKrafft2020}, \cite{KrafftVolokitin2024}) and based on the
two-dimensional (2D) modeling of electrostatic wave turbulence in a plasma
with a given spectrum of density fluctuations. In contrast to these studies,
we take into account here a weakly magnetized plasma with $\omega_{c}/\omega_{p}\lesssim0.2$ and study electromagnetic wave
radiation by upper-hybrid wave turbulence (\cite{Krafft2019}).

\begin{figure*}
    \centering
     \includegraphics[width=0.8\textwidth]{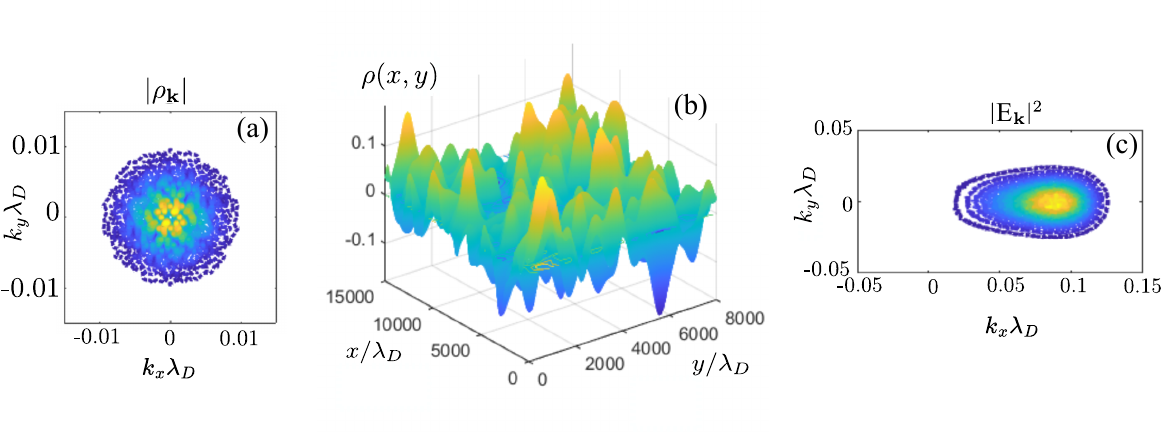}
    \caption{Example of distributions of upper-hybrid waves and density
fluctuations in the plasma source, at initial time $t=0$, for $\omega
_{c}/\omega_{p}=0.1.$ (a) Density fluctuations' spectrum $\left\vert
\rho_{\mathbf{k}}\right\vert =|\delta n_{\mathbf{k}}/n_{0}|$ in
the map $(k_{x}\lambda_{D},k_{y}\lambda_{D});$ $\delta n_\mathbf{k}$ is the Fourier component of $\delta n$. (b) Corresponding spatial distribution
$\rho(x,y)=\delta n(x,y)/n_{0}$ in the 2D map $(x/\lambda_{D},y/\lambda_{D}),$
with $\Delta N=0.05$.  (c) Electric
field energy spectrum $\left\vert \mathbf{E}_{\mathbf{k}}\right\vert ^{2}$ in
the map $(k_{x}\lambda_{D},k_{y}\lambda_{D})$, which
mimics the wave energy distribution generated by an electron
beam.{\scriptsize \ } All variables are normalized. The lengths of the simulation plane are
$L_{x}=15000\lambda_{D}$\ and $L_{y}=8000\lambda_{D}$, with $N_x=4096$ and $N_y=2048$ grid points.}
    \label{fig1}
\end{figure*}

Initially, a spectrum of density fluctuations  $\delta n$ with random phases is
set in the plasma with
\begin{equation}
\int_{2D}\frac{dxdy}{L_{x}L_{y}}\left(  \frac{\delta n}{n_{0}}\right)
^{2}=\left\langle \left(  {\delta n/n_{0}}\right)  ^{2}\right\rangle
=\left(  \Delta N\right)  ^{2},\label{DeltaN}
\end{equation}
where $L_x$ and $L_y$ are the lengths of the simulation plane,  and $\Delta N$ varies typically in the range
$0\lesssim\Delta N\lesssim0.05$. Density fluctuations $\delta n$ are assumed to present wavelengths much larger
than those of upper-hybrid waves.
For definiteness, we assume that their dynamics follows the linear equation
\begin{equation}
\left(  \frac{\partial^{2}}{\partial t^{2}}-c_{s}^{2}\Delta\right)
\frac{\delta n}{n_{0}}\simeq 0,
\label{is-eq}
\end{equation}
where $c_{s}=(  \left(  T_{e}+3T_{i}\right)  /m_{i})  ^{1/2}$ is
the ion acoustic velocity. Indeed, we neglect ponderomotive effects and thus
do not take into account nonlinear wave-wave interactions that are not
dominant processes in randomly inhomogeneous plasmas; interactions of electrostatic waves with density fluctuations and their subsequent transformations (refraction, reflection, scattering, tunneling, mode conversion) play the main role here. 
We consider upper-hybrid waves (also designated in  other of our works as Langmuir/$\mathcal{Z}$-mode or $\mathcal{LZ}$ waves) with the dispersion
\begin{equation}
\omega\simeq\omega_{p}+\frac{3}{2}\omega_{p}\left(  k\lambda_{D}\right)
^{2}+\frac{\omega_{c}^{2}}{2\omega_{p}}\frac{k_{{\perp}}^{2}}{k^{2}}\left(
1-\frac{\omega_{p}^{2}}{c^{2}k^{2}}\right)  , \label{dispersion}%
\end{equation}
where the condition $c^{2}k^{2}\gg\omega_{p}^{2}$ of negligible electromagnetic
contribution is supposed to be satisfied (see also Appendix $A$); $k_{\perp}$ is the perpendicular wavevector modulus; $\lambda_D$ id the electron Debye length. Then, the
slowly varying envelope $\tilde{\varphi}(\mathbf{r},t)$ of the upper-hybrid
potential $\varphi(\mathbf{r},t)$ evolves according to the high-frequency modified Zakharov equation 
including weak magnetic effects (\cite{KrasnoselskikhSotnikov1977})
\begin{equation}
\nabla^{2}\left(  i\frac{\partial\tilde{\varphi}}{\partial t}+\frac
{3\omega_{p}}{2}\lambda_{D}^{2}\nabla^{2}\tilde{\varphi}\right)  -\frac
{\omega_{c}^{2}}{2\omega_{p}}\nabla_{\perp}^{2}\tilde{\varphi}\simeq
\frac{\omega_{p}^{2}}{2\omega}\nabla\cdot\left(  \frac{\delta n}{n_{0}}%
\nabla\tilde{\varphi}\right)  ,\label{hf-eq}
\end{equation}
where $\partial\left\vert \delta n\right\vert /\partial t\ll\omega\left\vert
\delta n\right\vert $. In order to follow the dynamics of the wave potential $\tilde{\varphi}(\mathbf{r},t)$ and the ion density perturbation  $\delta n(\mathbf{r},t)$ (involving applied density fluctuations and induced ion perturbations), equations (\ref{is-eq}) and (\ref{hf-eq}) are solved
numerically in a 2D plane ($x,y$) of lengths $L_{x}$ and $L_{y}$,
respectively, where the ambient magnetic field $\mathbf{B}_{0}$ is directed along the $x$-axis.
Since we further assume that high-frequency waves have no back
influence on density fluctuations, the problem is linear in terms of wave
amplitudes, which are normalized below by the initial high-frequency
electrostatic wave energy
\begin{equation}
W_{UH}=\int_{2D}\frac{dxdy}{L_{x}L_{y}}\frac{\left\vert \nabla\tilde\varphi
(t=0)\right\vert ^{2}}{16\pi},\label{WL1}%
\end{equation}
which ranges typically as $10^{-6}\lesssim W_{UH}\lesssim10^{-4}$, so that ponderomotive
effects are negligible, as mentioned above (see also (\ref{is-eq})).
Figs. \ref{fig1}a,b show an example of initial density spectrum and spatial distribution, respectively. At $t=0$, the energy of high-frequency waves is set in the form of a drifted
Gaussian stretched along the magnetic field direction, in order to mimic wave
excitation by an electron beam (Fig. \ref{fig1}c); the waves' phases
are chosen random. 
We assume that electric current $ \mathbf{\delta j}(\mathbf{r},t)$ resulting from
the interactions of upper-hybrid waves with density fluctuations
$\delta n$ is relatively small, as $\Delta N$ is small. 
It can be calculated at each time $t$ and position $\mathbf{r}$ in the turbulent and randomly inhomogeneous source. Electromagnetic waves it radiates are supposed to propagate further away to infinity through a uniform plasma. For the reasons mentioned just above, we suppose
that the influence of plasma inhomogeneities on the propagation of
electromagnetic waves is negligible.

\begin{figure*}
    \centering
    \includegraphics[width=0.7\textwidth]{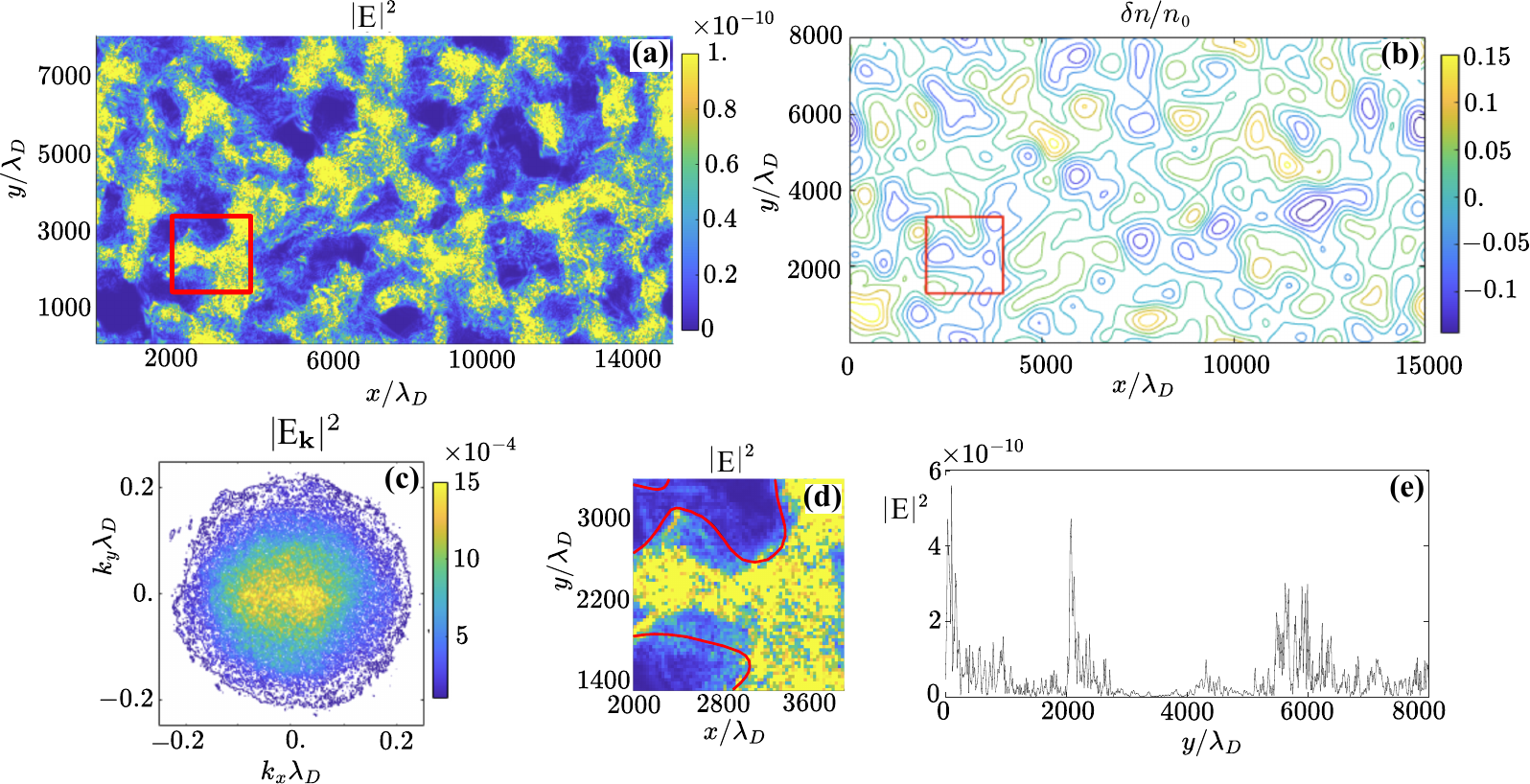} 
    \caption{Wave and density turbulence at  $\omega_{p}t\simeq7600.$ (a) Spatial
distribution of the upper-hybrid wave energy $\left\vert \mathbf{E}\right\vert
^{2}$ in the map $(x/\lambda_{D},y/\lambda_{D});$ the red square delimits the
region where a zoom is shown in (d). (b) Spatial
distribution of  $\delta n(x,y)/
n_{0}$ in the map $(x/\lambda_{D},y/\lambda_{D});$ the red square is
the same as in (a). (c)
Upper-hybrid wave spectrum $\left\vert \mathbf{E}_{\mathbf{k}}\right\vert
^{2}$  in the map $(k_{x}\lambda_{D},k_{y}\lambda_{D})$. (d) Zoom of the domain delimited by a red square
in (a); the isocontours with vanishing ion density perturbation $\delta
n_{i}=0$ are represented by red lines. (e) Profile along $y$, at fixed $x,$ of
the upper-hybrid wave energy $\left\vert \mathbf{E}\right\vert ^{2}$. Energies are shown in arbitrary units. Initial conditions
are shown in Fig. \ref{fig1}.}
    \label{fig2}
\end{figure*}
Fig. \ref{fig2} shows, after some time evolution, the distributions of the spatial high-frequency
wave energy (Figs. \ref{fig2}a,d) and of the ion density perturbation $\delta n_i(\mathbf{r},t)\simeq\delta n(\mathbf{r},t)$ (Fig. \ref{fig2}b), as well as the corresponding wave spectrum 
(Fig. \ref{fig2}c), obtained by solving equations (\ref{is-eq}) and (\ref{hf-eq}) with the initial
conditions of Fig. \ref{fig1}. The upper-hybrid spectrum broadens and tends
to become quasi-isotropic with time; localized wave packets are
formed in space (Figs. \ref{fig2}a,d) which are trapped in regions of reduced density. Wave energy profiles (Fig. \ref{fig2}e) show the formation of clumps of wavepackets.

The next sections are devoted to calculate the energy radiated by the plasma source at frequency $\omega_p$ in the three electromagnetic modes $\mathcal{O}$, $\mathcal{X}$ and $\mathcal{Z}$.

\subsection{Electromagnetic wave radiation by the source}

Three modes of high-frequency electromagnetic waves exist in a weakly magnetized
plasma at $\omega\simeq\omega_p$, which are the ordinary $\mathcal{O}$-mode (with the cutoff frequency
$\omega_{p}$), as well as the fast and slow extraordinary modes $\mathcal{X}$
and $\mathcal{Z}$, with cutoff frequencies 
$\omega_{+}=({\omega_{p}^{2}+\omega_{c}^{2}/4})^{1/2}+\omega
_{c}/2$  and  $\omega_{-}=({\omega_{p}^{2}+\omega_{c}^{2}/4})^{1/2}-\omega
_{c}/2$, respectively. If $\omega_{c}/\omega_{p}$ is
not too small as, for example, if $\omega_{c}/\omega_{p}\gtrsim\Delta N$,
these modes turn out to be separated in frequency at small
$k$, making it  possible to consider their emissions via 
linear mode conversion (LMC) at constant frequency of upper-hybrid waves
on plasma density fluctuations separately.

\subsubsection{Ordinary mode emission}
Let us first study $\mathcal{O}$-mode radiation in a weakly magnetized
and randomly inhomogeneous plasma. Starting from Maxwell equations, the dynamics of electromagnetic  radiation can be described by using
the wave magnetic field $\mathbf{B}$ (\cite{VolokitinKrafft2020})
\begin{equation}
\left(  \frac{\partial^{2}}{\partial t^{2}}-c^{2}\nabla^{2}\right)
\mathbf{B}=4\pi c\mathbf{\nabla}\times\left(  \delta\mathbf{J}+\delta
\mathbf{\mathbf{j}}\right)  \mathbf{,} \label{Maxwell}%
\end{equation}
where the current density $\delta\mathbf{J}=-en_{0}\mathbf{v}_{e}$ is due to
electrons moving with velocity $\mathbf{v}_{e}$, which is given
in linear approximation by%
\begin{equation}
\mathbf{v}_{e}\simeq-\frac{ie}{m_{e}\omega}\mathbf{E}_{\Vert}+\frac{ie}{m_{e}%
}\frac{\omega}{\omega_{c}^{2}-\omega^{2}}\left(  \mathbf{E}_{\perp}%
+i\frac{\omega_{c}}{\omega}\mathbf{h}\times\mathbf{E}_{\perp}\right)  ,
\label{ve1}%
\end{equation}
where $\mathbf{E}_{\Vert}$ and $\mathbf{E}_{\perp}$ are the parallel and
perpendicular electric fields of the upper-hybrid waves of frequency $\omega$;
$\mathbf{h}=\mathbf{B}_{0}/B_{0}$ is a unitary vector. The
electric fields $\mathbf{E}_{\Vert}=\delta\mathbf{E}_{\Vert}-\mathbf{\nabla
}_{\Vert}\varphi$ and $\mathbf{E}_{\perp}=\delta\mathbf{E}_{\perp
}-\mathbf{\nabla}_{\perp}\varphi$ contain small non-potential contributions
$\delta\mathbf{E}_{\perp}$ and $\delta\mathbf{E}_{\Vert}$, that can be
neglected here since (see Appendix $A$)
\begin{equation}
\left\vert \delta\mathbf{E|}/|\nabla\varphi\right\vert \simeq\frac{\omega
}{c^{2}k^{2}}\frac{\omega_{c}\omega_{p}^{2}}{\omega^{2}-\omega_{c}^{2}}%
\simeq\frac{\omega_{p}^{2}}{c^{2}k^{2}}\frac{\omega_{c}}{\omega}\label{l1}.
\end{equation}
The external current density $\delta\mathbf{j,}$ which results from 
interactions of upper-hybrid waves with much slower density fluctuations $\delta n$,
oscillates at a frequency close to $\omega_{p}$. As $\omega_c/\omega_p$ is weak, frequencies of upper-hybrid waves only
slightly differ from $\omega_{p}$ (equation (\ref{dispersion}) and Appendix $B$). This suggests that
frequencies of electromagnetic ordinary waves are also close to $\omega_{p}$.

The assumed weak interactions between electrostatic and electromagnetic waves via linear mode conversion (LMC) at constant frequency
allows us to isolate fast oscillating phases and to consider only the
evolution of the slowly varying envelopes $\mathbf{\tilde{B}}$ and
$\delta\mathbf{\mathbf{\tilde{j}}}$ of $\mathbf{B=}\operatorname{Re}%
(\mathbf{\tilde{B}}\left(  t\right)  e^{-i\omega_{p}t})$ and $\delta
\mathbf{j}=\operatorname{Re}(\delta\mathbf{\mathbf{\tilde{j}}}(t)e^{-i\omega
_{p}t})$, respectively. As the linear current $\delta\mathbf{J}$ can be
expressed through electric fields' amplitudes
\begin{equation}
4\pi\delta\mathbf{J}=\mathbf{-}4\pi en_{0}\mathbf{v}_{e}\simeq\frac
{i\omega_{p}^{2}}{\omega}\left(  \mathbf{E}+i\frac{\omega\omega_{c}}%
{\omega^{2}-\omega_{c}^{2}}\mathbf{h}\times\mathbf{E}+\frac{\omega_{c}^{2}%
}{\omega^{2}-\omega_{c}^{2}}\mathbf{E}_{\perp}\right)  ,\label{DeltaJ}%
\end{equation}
equation (\ref{Maxwell}) leads to
\begin{equation}
\left(  i\frac{\partial}{\partial t}-\frac{c^{2}}{2\omega_{p}}\hat{R}\right)
\mathbf{\tilde{B}}=4\pi c\mathbf{\nabla}\times\delta\mathbf{\mathbf{\tilde{j}%
},}%
\end{equation}
where we took into account that $\left\vert \partial^{2}\mathbf{\tilde{B}%
/}\partial t^{2}\right\vert \ll\omega_{p}\left\vert \partial\mathbf{\tilde{B}%
}/\partial t\right\vert $; $\hat{R}$ is a tensor operator involving all
magnetic effects, which cannot be expressed explicitly in an easy way.
However, using the Fourier components $\mathbf{B}_{\mathbf{k}\text{ }}$of
$\mathbf{\tilde{B}}=\sum_{\mathbf{k}}\mathbf{B}_{\mathbf{k}}\left(  t\right)
e^{i\mathbf{k\cdot r}}$, i.e. plane waves with polarization
vectors\ represented by $\mathbf{a}_{\mathbf{k}}=\mathbf{B}_{\mathbf{k}
}/B_{\mathbf{k}}$, where  $B_{\mathbf{k}}$ is the amplitude of $\mathbf{B}_{\mathbf{k}}$, we get
\begin{equation}
\left(  i\frac{\partial}{\partial t}-\Delta\omega_{\mathbf{k}}\right)
B_{\mathbf{k}}\simeq-\frac{2\pi c}{\omega_{p}}\mathbf{a}_{\mathbf{k}}^{\ast
}\cdot\left(  \mathbf{\nabla}\times\delta\mathbf{\mathbf{\tilde{j}}}\right)
_{\mathbf{k}}, \label{dBkdt}%
\end{equation}
where $\Delta\omega_{\mathbf{k}}=\omega_{\mathbf{k}}-\omega_{p}$ is the
frequency detuning and $\omega_{\mathbf{k}}=\omega(\mathbf{k})$ is the
dispersion relation. Of course, such a description is approximate and its
accuracy depends on the validity of the assumptions made; those will be
verified by our simulations, as shown below. The current density
$\delta\mathbf{j}=-e\delta n\mathbf{v}_{e} \label{cur1}$
is generated by the interactions (scattering) of upper-hybrid waves with the
slowly varying density fluctuations $\delta n$; keeping only terms
leading to electromagnetic wave radiation, we can write its slowly varying
envelope $\delta\mathbf{\mathbf{\tilde{j}}}$ as
\begin{equation}
4\pi\delta\mathbf{\mathbf{\tilde{j}}}\simeq\mathbf{-}\frac{i\omega_{p}%
^{3}}{\omega_p^{2}-\omega_{c}^{2}}\frac{\delta n}{n_{0}}\left(
\mathbf{\nabla}\tilde{\varphi}+i\frac{\omega_{c}}{\omega}\mathbf{h}%
\times\mathbf{\nabla}\tilde{\varphi}-\frac{\omega_{c}^{2}}{\omega^{2}%
}\mathbf{h}\left(  \mathbf{h\cdot\nabla}\right)  \tilde{\varphi}\right),
\label{Current}%
\end{equation}
where $\omega\simeq\omega_p$. Note that, compared to the unmagnetized plasma case, two new terms involving $\omega_{c}$ have appeared. For future use, let us define the vector
$\mathbf{G}$ as
\begin{equation}
\mathbf{G}=4\pi \frac{(\omega_p^{2}-\omega_{c}^{2})}{\omega_{p}^{3}
}\mathbf{\nabla}\times\delta\mathbf{\tilde{j}.} \label{G}%
\end{equation}
Then, the growth with time of waves' magnetic and electric field energies can
be determined by theoretical and numerical integration of (\ref{dBkdt}). To
fulfill this task, the current \ $\delta\mathbf{\mathbf{\tilde{j}}}(x,y,t)$
(\ref{Current}) is calculated using the potentials $\tilde{\varphi}(x,y,t)$
and the density modulation $\delta n(x,y,t)$ provided at each time $t$ and
position $(x,y)$ by solving jointly equations (\ref{is-eq}) and (\ref{hf-eq}).
In 2D geometry, the $x$-axis ($y$-axis) is directed along (across)
$\mathbf{B}_{0}$, and the $z$-axis (with unitary vector $\mathbf{z}$) is
perpendicular to the simulation map $(x,y)$, with $\partial/\partial
z=\partial_{z}=0.$ According to the linear analysis of $\mathcal{O}$-mode
dispersion and polarization in a weakly magnetized plasma (see Appendix $B$), its magnetic field component $B_{z}$ is dominant ($\left\vert
B_{x,y}\right\vert \ll\left\vert B_{z}\right\vert $) so that $\mathbf{B}\simeq B_{z}\mathbf{z.}$ Moreover, the corresponding
wave electric field lies into the ($x,y$) plane and $E_{z}\simeq0$. These
conditions can be satisfied for a vector potential of the form $\mathbf{A=(}%
A_{x},A_{y},0),$ with $B_{z}={\partial A_{y}}/{\partial x}-{\partial A_{x}}/{\partial
y}$,  $E_{x,y}=-{\partial A_{x,y}}/{c\partial t} \label{Bz}$ and $\left(  \mathbf{\nabla\times B}\right)  _{z}=0$. The small
 neglected field components $E_{z}$, $B_{x}$ and $B_{y}$ have nevertheless
to be taken into account when determining the wave dispersion (see Appendix $B$).
According to the above approximations, we can write that $\mathbf{a}%
_{\mathbf{k}}\simeq\mathbf{z}$ and present the equation (\ref{dBkdt}) in the form%
\begin{equation}
\left(  i\frac{\partial}{\partial t}+\omega_{p}-\omega_{\mathbf{k}}\right)
B_{z\mathbf{k}}\simeq-\frac{c}{2}\frac{\omega_{p}^2}{\omega_p^{2}-\omega
_{c}^{2}}G_{z\mathbf{k}}, \label{dBzkdt}%
\end{equation}
where $B_{z\mathbf{k}}$ and $G_{z\mathbf{k}}$ are the Fourier components of
the envelopes of $B_{z}$ and $G_{z}=\mathbf{G}\cdot\mathbf{z}$ (\ref{G})
\begin{equation}
G_{z}=-i\left(  \mathbf{\nabla}\frac{\delta n}{n_{0}}\times\mathbf{\nabla}%
\tilde{\varphi}\right)  _{z}-i\frac{\omega_{c}^{2}}{\omega^{2}}\left(
\mathbf{h}\times\mathbf{\nabla}\right)  _{z}\left(  \frac{\delta n}{n_{0}%
}\mathbf{\partial}_{\parallel}\tilde{\varphi}\right)  , \label{GZ}%
\end{equation}
respectively, and $\mathbf{\partial}_{\parallel}=\mathbf{\partial}_{x}$. We
took into account in equation (\ref{GZ}) that $\mathbf{h}\cdot\mathbf{z}=0$ and
$\partial_{z}=0$. Other components of $\mathbf{G}$ provide only small
contributions to $\mathcal{O}$-mode radiation, but they are essential
regarding the $\mathcal{X}$- and $\mathcal{Z}$-modes' generation, as shown
hereafter. One observes in equations (\ref{dBzkdt})-(\ref{GZ}) that the only
contribution of magnetic effects to $G_{z}$ is of the second order in
$\omega_{c}$, and can therefore be neglected in a solar wind plasma.

Simulation results obtained by solving (\ref{dBzkdt})-(\ref{GZ}) are presented below in dimensionless variables according to
the normalization $\omega_{p}t\rightarrow t$, $\mathbf{r/}\lambda_{D}\rightarrow\mathbf{r}$, $\mathbf{k}\lambda_{D}\rightarrow\mathbf{k}$ and
$\mathbf{\tilde{E}}/\sqrt{16\pi W_{UH}}\mathbf{\rightarrow E}(E_{x},E_{y})$
with $\int_{2D}({dxdy}/{L_{x}L_{y}})({|\mathbf{\tilde{E}|}^{2}}/{16\pi W_{UH}})=1$, where $\mathbf{\tilde{E}=-\nabla}\tilde{\varphi}$ is the electric field
envelope; the dimensionless spectral magnetic field is defined as $b_{\mathbf{k}}\left(  t\right)  ={B_{z\mathbf{k}}\left(t\right)
}/{\sqrt{16\pi W_{UH}}}.$
Then,  equation (\ref{dBzkdt}) can be written in dimensionless form as
\begin{equation}
\left(  i\frac{\partial}{\partial t}-\Delta\omega_{\mathbf{k}}\right)
b_{\mathbf{k}}(t)\simeq-\frac{c_{L}}{2}\frac{\omega_{p}^{2}}{\omega_{p}%
^{2}-\omega_{c}^{2}}\hat{G}_{z\mathbf{k}}, \label{dbkdt}%
\end{equation}
where $\Delta\omega_{\mathbf{k}}=(\omega_{\mathbf{k}}-\omega
_{p})  /\omega_{p}$ (note that the same notation $\Delta\omega
_{\mathbf{k}}$\ is used for both the normalized equation (\ref{dbkdt}) and the
physical one\ (\ref{dBzkdt})), $c_{L}=c/v_{T}$ and

\begin{equation}
\hat{G}_{z\mathbf{k}}=G_{z\mathbf{k}}\frac{\lambda_{D}}{\sqrt{16\pi W_{UH}}%
}\simeq -i \left(\frac{\partial\tilde{\varphi}}{\partial y}\frac{\partial}{\partial
x}\frac{\delta n}{n_{0}}-\frac{\partial\tilde{\varphi}}{\partial x}%
\frac{\partial}{\partial y}\frac{\delta n}{n_{0}} 
+ \frac {\omega_{c}^{2}}
{\omega_p^{2}} \frac{\partial}{\partial y}\left(  \frac{\delta n}{n_{0}}%
\frac{\partial\tilde{\varphi}}{\partial x}\right)  \right)  _{\mathbf{k}}.
\label{gzk}%
\end{equation}

Equation (\ref{dbkdt}) can be integrated as
\begin{equation}
b_{\mathbf{k}}\left(  t\right)  \simeq\frac{ic_{L}}{2}\frac{\omega_{p}^{2}%
}{\omega_{p}^{2}-\omega_{c}^{2}}\int_{0}^{t}\hat{G}_{z\mathbf{k}}\left(
t^{\prime}\right)  e^{i\Delta\omega_{\mathbf{k}}\left(  t^{\prime}-t\right)
}dt^{\prime}, \label{bk-int}%
\end{equation}

so that $b_{\mathbf{k}}\left(  t+\Delta t\right)  $ can be expressed as a function of
$b_{\mathbf{k}}\left(  t\right)  $ owing to the explicit\ numerical scheme
used in our previous works (\cite{VolokitinKrafft2020}).
The normalized magnetic and electric energies $\mu\left(  t\right)  $ and
$\eta(t)$ inside the volume $L_{x}L_{y}$ can be expressed as
\begin{equation}
\mu\left(  t\right)  =\sum_{\mathbf{k}}\left\vert b_{\mathbf{k}}\right\vert
^{2},\text{ \ \ \ \ \ \ }\eta(t)=\sum_{\mathbf{k}}\frac{\left\vert
b_{\mathbf{k}}\right\vert ^{2}}{c_{L}^{2}\mathbf{k}^{2}}.\label{WB}%
\end{equation}
The corresponding radiation rates are $\dot{\mu}=d\mu/dt$ and $\dot{\eta}=d\eta/dt$. Note that
$\mu\left(  t\right)  \ll\eta(t),$ so that $\eta(t)$ can also be considered as the total energy carried by the electromagnetic ordinary waves. These quantities depend
on $c_{L},$ $\omega_{c}/\omega_{p}$, and $\Delta N$, i.e. on the plasma
electron temperature, magnetization and average level of density
inhomogeneities, as well as on the initial upper-hybrid waves' and density
fluctuations' spectra. \emph{\ }One can expect that for $\Delta N\gtrsim
0.01,$\ the upper-hybrid wave energy spectrum tends to izotropize (\cite{KrafftVolokitin2021}) due to wave transformations on density fluctuations occurring at a fast rate, generally exceeding those of other processes, as wave attenuation. Then it becomes quasi-isotropic asymptotically, as  expected in the solar wind. Since only $\dot{\eta}$ has a
clear physical meaning and is directly related to the radiation intensity, we
focus hereafter on its dependence on $c_{L}$, $\omega_{c}/\omega_{p}$, and
$\Delta N$, assuming that possible variations of the upper-hybrid energy
spectra do not significantly affect the growth rate of electromagnetic waves
radiated by a given volume of turbulent plasma.

\begin{figure}
    \centering
    \includegraphics[width=0.6\textwidth]{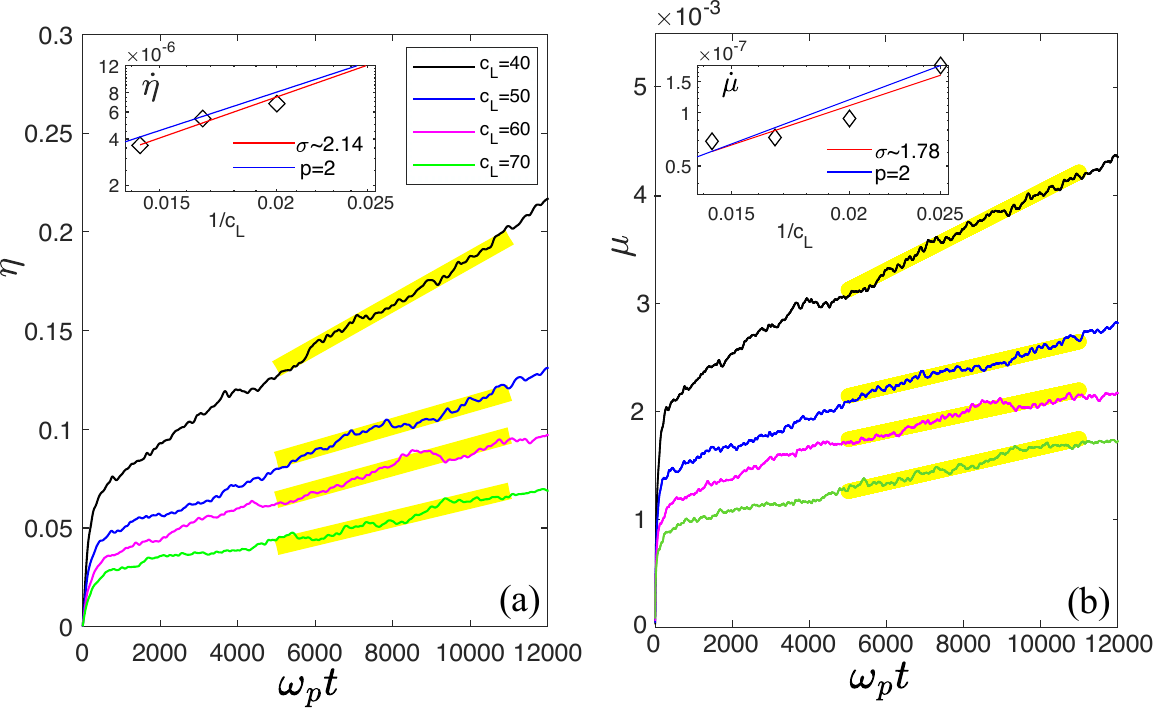}
    \caption{Time variations of the electric (a) and magnetic (b) wave energies
$\eta(t)$ and $\mu(t)$, respectively, in an unmagnetized plasma$\ (\omega
_{c}=0),$ for $\Delta N=0.03$ and four values of $c_{L}$ (see legend in (a)). The insets show, in logarithmic scales, the variations of the corresponding radiation rates $\dot{\eta}$ and $\dot{\mu}$ as a function of $1/c_{L}$, which exhibit scaling indices $\sigma\simeq2.14$ (a)
and $\sigma\simeq1.78$ (b), respectively, to be compared with the value $p=2$ (\ref{eta1}) and the blue lines; the values of $\dot{\eta}$ and
$\dot{\mu}$ provided by the simulations are indicated by black diamonds. The
superimposed thick yellow lines represent the linear interpolations
of $\eta(t)$ and $\mu(t)$ at large times, that provide the radiation rates. All variables are normalized.}
    \label{fig3}
\end{figure}

Let us first present the results obtained for an unmagnetized plasma source, when the only
electromagnetic mode is the ordinary one. Equation
(\ref{dbkdt}) can be applied to this case, with $\omega_{c}=0$\ and
electromagnetic wave dispersion $\omega_{\mathbf{k}}\simeq\omega_{p}%
+c^{2}k^{2}/2\omega_{p}$. Figs. \ref{fig3}a-b show, for different
$c_{L},$ the time variations of electromagnetic and magnetic wave energies
$\eta\left(  t\right)  $ and $\mu\left(  t\right)  $, respectively, which grow linearly at asymptotic times. Their slopes, which represent the radiation rates
$\dot{\eta}$ and $\dot{\mu}$, exhibit small fluctuations, which can be
attributed to the statistical nature of wave turbulence. On average, $\dot{\eta}$
and $\dot{\mu}$ do not depend on time,  which is consistent with the analytical calculations
 presented in section 3 below. The insets in Figs. \ref{fig3}%
a-b show the dependence of the radiation rates $\dot{\eta}$ and $\dot{\mu}$ on $1/c_{L}$,
which convincingly demonstrates that the following power law is satisfied with good accuracy $
\dot{\eta}\propto1/c_{L}^{2}, \label{eta1} $
as the scaling indices calculated by interpolating the points provided by the
numerical simulations are $\sigma\simeq2.14$ (Fig. \ref{fig3}a) and $\sigma\simeq1.78$ (Fig. \ref{fig3}b), respectively. The discrepancy between $\dot{\eta
}$ and $\dot{\mu}$ is due to  numerical features inherent to our modeling,
but also to differences between time variations of spectral electric and
magnetic energies of electromagnetic waves generated during linear
transformations of upper-hybrid waves on density fluctuations. Note that such
process is only possible if the frequency detuning $\omega-\omega_{p}\sim
c^{2}k^{2}/2\omega_{p}$\ of the produced electromagnetic waves does not exceed
the spectral width of the scattered electrostatic waves, which can be
estimated as $\omega_{p}\Delta N$.%

\begin{figure}
    \centering
    \includegraphics[width=0.6\textwidth]{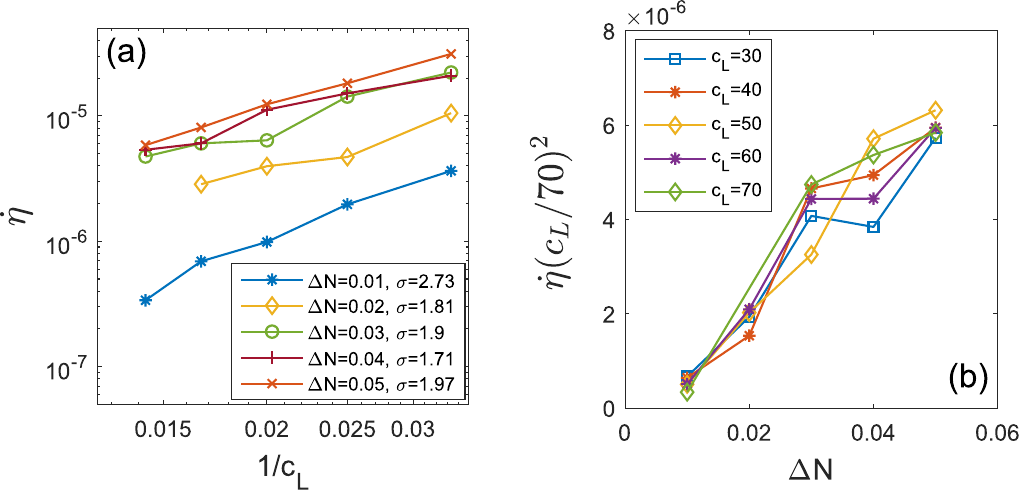}
    \caption{Electromagnetic $\mathcal{O}$-mode in an unmagnetized plasma.\emph{\ }(a) Variations of the radiation rate $\dot{\eta}$ as a function of
$1/c_{L}$, for different values of $\Delta N$, corresponding to the scaling
indices $\sigma$ listed in the  legend. (b) Variations of $\dot{\eta}(c_{L}/70)^{2}$ as
a function of $\Delta N$, for different values of $c_{L}\ $(see legend). The
initial density fluctuations and upper-hybrid wave turbulence spectra are both
anisotropic. (a) : logarithmic scales; (b) : linear scales. All variables are normalized.}
    \label{fig4}
\end{figure}

\begin{figure}
    \centering
    \includegraphics[width=0.6\textwidth]{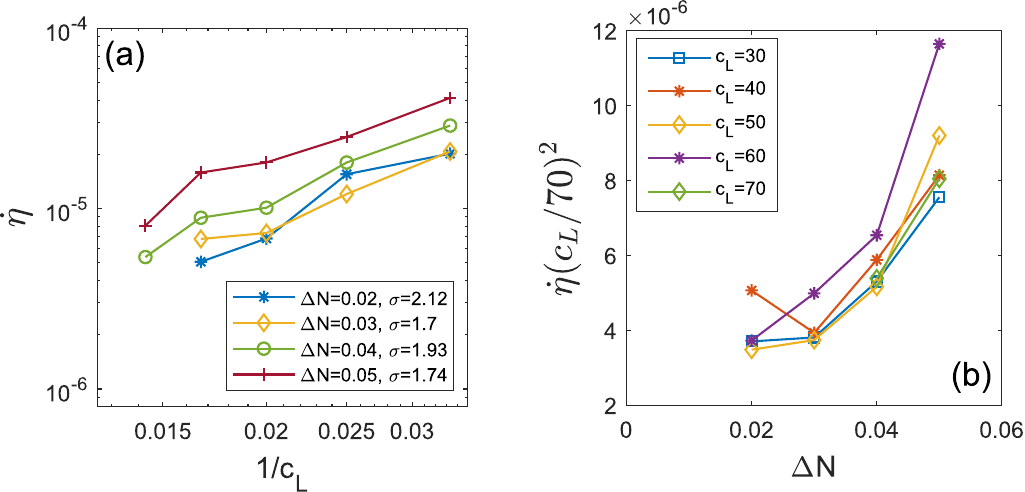}
    \caption{Electromagnetic $\mathcal{O}$-mode in an unmagnetized plasma.\emph{\ }(a) Variations of the radiation rate $\dot{\eta}$ as a function of
$1/c_{L}$, for different values of $\Delta N$, corresponding to the scaling
indices $\sigma$ listed in the  legend. (b) Variations of $\dot{\eta}(c_{L}/70)^{2}$ as
a function of $\Delta N$, for different values of $c_{L}\ $(see legend). The
initial density fluctuations and upper-hybrid wave turbulence spectra are isotropic and anisotropic, respectively. (a) : logarithmic scales; (b) : linear scales. All variables are normalized.}
    \label{fig5}
\end{figure}

\begin{figure}
    \centering
    \includegraphics[width=0.6\textwidth]{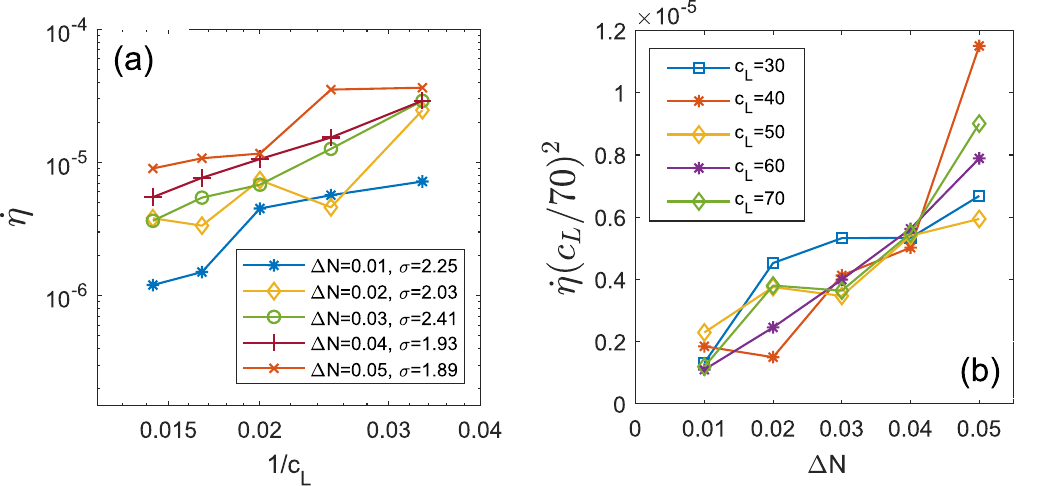}
    \caption{Electromagnetic $\mathcal{O}$-mode in an unmagnetized plasma.\emph{\ }(a) Variations of the radiation rate $\dot{\eta}$ as a function of
$1/c_{L}$, for different values of $\Delta N$, corresponding to the scaling
indices $\sigma$ listed in the  legend. (b) Variations of $\dot{\eta}(c_{L}/70)^{2}$ as
a function of $\Delta N$, for different values of $c_{L}\ $(see legend). The
initial density fluctuations and upper-hybrid wave turbulence spectra are anisotropic and isotropic, respectively. (a) : logarithmic scales; (b) : linear scales. All variables are normalized.}
    \label{fig6}
\end{figure}

\begin{figure}
    \centering
    \includegraphics[width=0.3\textwidth]{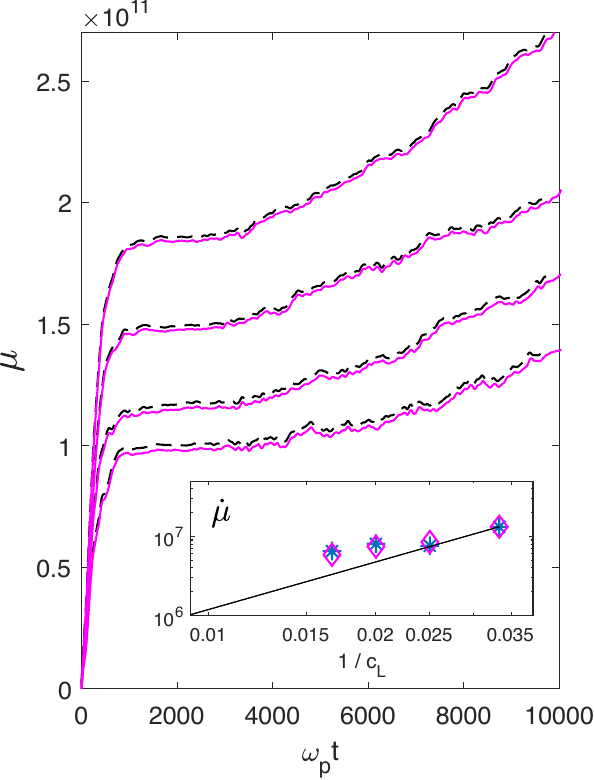}
    \caption{Time variations of the magnetic wave energy $\mu(t)$ in an
unmagnetized plasma$\ $($\omega_{c}=0$), for $\Delta N=0.03$ and $c_{L}=30,40,50$ and $60.$ The dashed black and solid  pink lines correspond
to simulations performed with the time steps $\omega_p\Delta t=1$ and $\omega_p\Delta t=5.4$,
respectively. The inset shows the radiation rates $\dot{\mu}$
as a function of $1/c_{L}$ (in logarithmic scale), estimated using the slopes of $\mu(t)$ at large
times, for the four values of $c_{L}$, and represented by blue stars ($\omega_p\Delta t=1$) and pink
diamonds ($\omega_p\Delta t=5.4$). For comparison, the solid black line corresponds to the
scaling index $p=2$. All variables are normalized.}
    \label{fig7}
\end{figure}

Fig. \ref{fig4} shows the variations of the electromagnetic ordinary waves' radiation rates
$\dot{\eta}$ as a function of $1/c_{L}$ (for $0.01\leq\Delta
N\leq0.05$) and of $\dot{\eta}(c_{L}/70)^{2}$ with $\Delta N$ (for different
$c_{L}$), for initial anisotropic density and upper-hybrid wave spectra.
Despite the observable scattering of points, the tendencies $\dot{\eta}%
\propto1/c_{L}^{\sigma}$ and $\dot{\eta}\propto\Delta N$ appear clearly. The numerically determined indices $\sigma$ actually form a distribution centered around $2.02$ (Fig. \ref{fig4}, left).
Then the ordinary mode radiation rate $\dot{\eta}$ in an unmagnetized plasma increases linearly with the
average level of density fluctuations $\Delta N$, whereas it depends on the
velocity ratio $v_{T}/c$ according to the power law $\dot{\eta}\propto\left(
v_{T}/c\right)  ^{\sigma}$, where $\sigma\sim2$. The same conclusions can  be stated when, initially, the density spectrum is isotropic
and the wave spectrum is anisotropic (Fig. \ref{fig5}), or inversely
(Fig. \ref{fig6}). Considering Figs. \ref{fig4},\ref{fig5} and
\ref{fig6}, one observes that\emph{\ }$\dot{\eta}$ typically extends
within the range $5 \cdot 10^{-7}\lesssim\dot{\eta}\lesssim 5 \cdot 10^{-5}$ and depends
significantly on the isotropy or anisotropy of initial wave and density spectra. 
The precision obtained for $\dot{\mu}$ and $\dot{\eta}$ depends on the
integration time step $\Delta t$ of the fast
numerical scheme used (\cite{VolokitinKrafft2020}) and on the time interval $\Delta T$ over which the linear
approximations of the asymptotic variations of $\mu\left(  t\right)  $ and
$\eta\left(  t\right)  $ are performed. Fig. \ref{fig7} shows the time
variations of $\mu\left(  t\right)  $ for simulations performed with $\omega_p\Delta
t=5.4$ and $\omega_p\Delta t=1$, showing small differences between both cases. Scattering of points can be attributed to the stochastic nature of the main
process at work, which is additionally forced by the finite number of waves
used in simulations. This induces unavoidable uncertainties in the
numerical interpolations of $\mu\left(  t\right)  $ and $\eta\left(  t\right)
$ at asymptotic times, that provide the radiation rates $\dot{\mu}$ and
$\dot{\eta}$, respectively. However, one finds close values of  $\dot{\mu}$ for both cases. Note that conditions  $\omega_p\Delta t< 5.4$ and  $\omega_p\Delta T\lesssim 10000$ have been used in this work.
The simulation results on $\mathcal{O}$-mode wave radiation in a weakly magnetized
plasma are discussed in the next section, together with those of extraordinary
modes' radiation, for comparison purposes.

\subsubsection{Extraordinary mode emission}   
Let us now consider electromagnetic waves in the $\mathcal{X}$- and $\mathcal{Z}$-modes 
and determine the equations that describe their radiation, as done above for the
$\mathcal{O}$-mode, assuming that the small ratio $\omega_{c}/\omega_p \leq 0.2$ is
sufficiently large for these modes to be excited independently of each other. Otherwise, for very weak ambient magnetic fields, our approach would require
further clarification, but in such
cases the role of plasma magnetization is insignificant in the presence of
density fluctuations. To achieve this goal, it seems a
priori possible to replace $\omega_{p}$ with the cutoff frequencies
$\omega_{\pm}\simeq (\omega_p^2+\omega_c^2/4)^{1/2} \pm \omega_c/2$ in the frequency detuning $\Delta
\omega_{\mathbf{k}}=\omega_{\mathbf{k}}-\omega_{p}$ (\ref{dBzkdt}) but
this is not exact, mainly because $\mathcal{X}$-  and $\mathcal{Z}$-modes have
different polarizations. 

In 2D geometry, the electric field of $\mathcal{X}$-
and $\mathcal{Z}$-modes is perpendicular to $\mathbf{B}_{0}=B_0\mathbf{x}$ and the wave magnetic field $\mathbf{B=(}B_{x}%
,B_{y},B_{z})$ presents three non vanishing components. However, the 2D
condition $\partial/\partial z=0$ allows us to propose a description using a
vector potential $\mathbf{A=(}0,A_{y},A_{z})$ with two perpendicular
components only, with $B_{x}={\partial A_{z}}/{\partial y}$,  $B_{y}=-{\partial A_{z}
}/{\partial x}$, and $B_{z}={\partial A_{y}}/{\partial x}$,
showing that the divergence-free condition $\mathbf{\nabla\cdot B}=0$ is
fulfilled. Accordingly, we have $E_{x}\simeq0$, $E_{y}=-{\partial A_{y}}/{c\partial t},$ and $E_{z}
=-{\partial A_{z}}/{c\partial t}$.
Using the Maxwell-Faraday law in the form $\mathbf{B}\simeq\left(
c/i\omega\right)  \mathbf{\nabla\times E}$, equation (\ref{Maxwell}) can be
written as
\begin{equation}
\mathbf{\nabla\times}\left[  \left(  \frac{\partial^{2}}{\partial t^{2}%
}+\omega_{p}^{2}-c^{2}\nabla^{2}\right)  \mathbf{E}+i\frac{\omega_{p}%
^{2}\omega_{c}\omega}{\omega^{2}-\omega_{c}^{2}}\left(  \mathbf{h}%
\times\mathbf{E}-i\frac{\omega_{c}}{\omega}\mathbf{E}_{\perp}\right)  \right]
\simeq4\pi i\omega\mathbf{\nabla}\times\delta\mathbf{j,}\label{rot}%
\end{equation}
where we took into account that $\mathbf{\nabla}\times\nabla^{2}%
\mathbf{E=\nabla^{2}}\left(  \mathbf{\nabla\times E}\right)  $
($\mathbf{\nabla}^{2}$ is here the Laplacian vector). Then, integrating
(\ref{rot}), we obtain the following system for $E_{y}$ and  $E_{z}$
\begin{equation}
\left(  \frac{\partial^{2}}{\partial t^{2}}+\frac{\omega_{p}^{2}\omega^{2}%
}{\omega^{2}-\omega_{c}^{2}}-c^{2}\nabla^{2}\right)  E_{y}-i\frac{\omega
_{p}^{2}\omega\omega_{c}}{\omega^{2}-\omega_{c}^{2}}E_{z}+\frac{\partial\Psi
}{\partial y}\simeq4\pi i\omega j_{y},\label{Ey1}%
\end{equation}
\begin{equation}
\left(  \frac{\partial^{2}}{\partial t^{2}}+\frac{\omega_{p}^{2}\omega^{2}%
}{\omega^{2}-\omega_{c}^{2}}-c^{2}\nabla^{2}\right)  E_{z}+i\frac{\omega
_{p}^{2}\omega\omega_{c}}{\omega^{2}-\omega_{c}^{2}}E_{y}\simeq4\pi i\omega
j_{z},\label{Ez}%
\end{equation}
where $j_{x}$, $j_{y}$, and $j_{z}$ are the components of $\delta\mathbf{j}$.
The term $\partial\Psi/\partial y$ ($\partial\Psi/\partial z=0,$ due to
2D geometry), which appears as a consequence of integration on the curl
operator $\nabla\times$\ (\ref{rot}), can be determined from general heuristic
considerations based on the fact that equation (\ref{Ey1}) results from
Maxwell equations, which are linear when the external current is vanishing,
and where electromagnetic wave fields appear only under differential
operators. This indicates that $\Psi$ only depends on the electric field
$\mathbf{E}$,\textbf{ }and linearly. In addition, $\Psi$\ must be a scalar
quantity. All this allows us to assume that\textbf{\ }$\Psi=c^{2}\left(
\nabla\cdot\mathbf{E}\right)  $, where the coefficient $c^{2}$ results from dimensional and dispersive considerations (Appendix $B$). In
2D geometry we get that $\Psi=c^{2}\left(  \partial E_{x}/\partial x+\partial
E_{y}/\partial y\right)  $; for transverse $\mathcal{X}$  and $\mathcal{Z}$ 
mode wave propagation, we must take into account the parallel electric field
$E_{x}$. As discussed in Appendix $B$, this component introduces some
corrections to wave dispersion, which are however small near the cutoff
frequencies, so that we can assume that ${\partial\Psi}/{\partial y}\simeq c^{2}{\partial^{2}E_{y}}/{\partial y^{2}}.$
Then equation (\ref{Ey1}) becomes
\begin{equation}
\left(  \frac{\partial^{2}}{\partial t^{2}}+\frac{\omega_{p}^{2}\omega^{2}%
}{\omega^{2}-\omega_{c}^{2}}-c^{2}\nabla^{2}+c^{2}\frac{\partial^{2}}{\partial
y^{2}}\right)  E_{y}-i\frac{\omega_{p}^{2}\omega\omega_{c}}{\omega^{2}%
-\omega_{c}^{2}}E_{z}\simeq4\pi i\omega j_{y}.\label{Ey2}%
\end{equation}
Introducing in equations (\ref{Ez})-(\ref{Ey2}) the notations $E_{\pm}%
=E_{z}\pm iE_{y}$ and $j_{\pm}=j_{z}\pm ij_{y},$ we get\ after
straightforward calculations that 
\begin{equation}
\left(  \frac{\partial^{2}}{\partial t^{2}}+\frac{\omega_{p}^{2}\omega^{2}%
}{\omega^{2}-\omega_{c}^{2}}\left(  1\pm\frac{\omega_{c}}{\omega}\right)
-c^{2}\nabla^{2}+\frac{c^{2}}{2}\frac{\partial^{2}}{\partial y^{2}}\right)
E_{\pm}\simeq4\pi i\omega j_{\pm},\label{EXz}%
\end{equation}
where $E_{+}$ and $E_{-}$ correspond to the fields of $\mathcal{X}$- and
$\mathcal{Z}$-mode waves, respectively; we have separated these modes, i.e. we have
set that $\partial_{y}^{2}E_{-}\rightarrow0$ ($\partial_{y}^{2}E_{+}%
\rightarrow0$) for the $\mathcal{X}$-mode ($\mathcal{Z}$-mode). Note that at
$\nabla^{2}=0$ ($k=0$) and without external currents ($\delta\mathbf{j}=0$),
equation (\ref{EXz}) predicts the existence of two kinds of oscillations that
satisfy $\omega\left(  \omega\mp\omega_{c}\right)  =\omega_{p}^{2},$ whose
solutions are the cutoff frequencies $\omega_{\pm}$, as it
should be at $k\simeq0$. Then, using $E_{\pm}=\operatorname{Re}(\tilde{E}%
_{\pm}\left(  t\right)  e^{-i\omega_{p}t})$ and $j_{\pm}=\operatorname{Re}%
\left(  \tilde{j}_{\pm}\left(  t\right)  e^{-i\omega_{p}t}\right)  $ to
separate the slow evolution of the envelopes $\tilde{E}_{\pm}$ and $\tilde
{j}_{\pm}$ from their fast phases' oscillations at $\omega_{p},$ and assuming
that $E_{+}$\ and $E_{-}$\ are sufficiently far apart in frequency to be
considered separately\emph{,} equation (\ref{EXz}) can be expressed as follows

\begin{equation}
\left(  -2i\omega_{p}\frac{\partial}{\partial t}-\omega_{p}^{2}+\frac
{\omega_{p}^{2}\omega^{2}}{\omega^{2}-\omega_{c}^{2}}\left(  1\pm\frac
{\omega_{c}}{\omega}\right)  -c^{2}\nabla^{2}+\frac{c^{2}}{2}\frac
{\partial^{2}}{\partial y^{2}}\right)  \tilde{E}_{\pm}\\ 
 \simeq 4\pi i\omega
\tilde{j}_{\pm}.\label{envelopes}
\end{equation}

Applying the Fourier transform $E_{\mathbf{k}}^{\pm}=\int_{2D}\tilde{E}_{\pm
}e^{-i\mathbf{k}\cdot\mathbf{r}}(dxdy/L_{x}L_{y}) $, we finally get the compact equation
\begin{equation}
\left(  i\frac{\partial}{\partial t}+\omega_{p}-\omega_{\mathbf{k}}^{\pm
}\right)  E_{\mathbf{k}}^{\pm}\simeq-2\pi ij_{\mathbf{k}}^{\pm},\label{XZEkdt}%
\end{equation}
where $j_{\mathbf{k}}^{\pm}$ ($E_{\mathbf{k}}^{\pm}$) is the Fourier transform
of $\tilde{j}_{\pm}$ ($\tilde{E}_{\pm}$) and
\begin{equation}
\omega_{\mathbf{k}}^{\pm}\simeq\omega_{p}\pm\frac{\omega_{c}}{2}+\frac
{c^{2}(k^{2}+k_{\parallel}^{2})}{4\omega_{p}}\simeq\omega_{\pm}+\frac
{c^{2}(k^{2}+k_{\parallel}^{2})}{4\omega_{p}},\label{disp}%
\end{equation}
where $k^{2}=k_{\parallel}^{2}+k_{\perp}^{2}.$ Note that we recover the same dispersion relations of $\mathcal{X}$- and
$\mathcal{Z}$-modes as found  in Appendix $B$ (but without the very small term proportional to
$\omega_{c}$) , where they are determined near the cutoff frequencies
$\omega_{\pm}$\ by using linear wave theory in $\mathbf{k}$-space. The external
current $\tilde{j}_{\pm}=\tilde{j}_{z}\pm i\tilde{j}_{y}$ results from
interactions between upper-hybrid fields and density
fluctuations; using equation (\ref{Current}) and neglecting the small second order term
in $\omega_{c}^{2}$, we get that
\begin{equation}
4\pi\delta\mathbf{\tilde{j}}=4\pi(\tilde{j}_{x},\tilde{j}_{y},\tilde{j}%
_{z})\simeq -\frac{i\omega_{p}^{2}\omega}{\omega^{2}-\omega_{c}^{2}}\frac{\delta
n}{n_{0}}\left(  \frac{\partial\tilde{\varphi}}{\partial x},\text{ }%
\frac{\partial\tilde{\varphi}}{\partial y},\text{ }i\frac{\omega_{c}}{\omega
}\frac{\partial\tilde{\varphi}}{\partial y}\right)  .\label{j-env}%
\end{equation}
Note that the parallel current $\tilde{j}_{x}$ can, under suitable
conditions, contribute to $\mathcal{O}$-mode radiation, but this case is not
considered here. Then we get
\begin{equation}
4\pi\tilde{j}_{\pm}=4\pi\left(  \tilde{j}_{z}\pm i\tilde{j}_{y}\right)
\simeq-\frac{\omega_{p}^{2}}{\left(  \omega_{c}\mp\omega\right)  }\frac{\delta
n}{n_{0}}\frac{\partial\tilde{\varphi}}{\partial y}.\label{jpm}%
\end{equation}
Introducing the normalized field amplitudes $e_{\mathbf{k}}^{\pm}\left(  t\right)  ={E_{\mathbf{k}}^{\pm}\left(
t\right)  }/{\sqrt{8\pi W_{UH}}}$ of
$\mathcal{X}$-$\ $and $\mathcal{Z}$-modes, we obtain the dimensionless form of equation (\ref{XZEkdt})
\begin{equation}
\left(  i\frac{\partial}{\partial t}-\Delta\omega_{\pm}\right)  e_{\mathbf{k}%
}^{\pm}\left(  t\right)  =iq_{\mathbf{k}}^{\pm}\left(  t\right)
,\label{xznorm}%
\end{equation}
with
\begin{equation}
\Delta\omega_{\pm}=\frac{\omega_{\mathbf{k}}^{\pm}-\omega_{p}}{\omega_{p}%
},\text{ \ \ \ \ \ \ \ \ \ }q_{\mathbf{k}}^{\pm}\left(  t\right)
=\frac{\omega_{p}}{2\left(  \omega_{c}\mp\omega\right)  }\left(  \frac{\delta
n}{n_{0}}\frac{\partial\tilde{\varphi}}{\partial y}\right)  _{\mathbf{k}%
},\label{xznorm2}%
\end{equation}
where $\partial\tilde{\varphi}/\partial y$ is normalized as indicated above. The density fluctuations $\delta n$ as well as the potentials
$\tilde{\varphi}$ and their derivatives are provided by the 2D modeling (see
Section 2.1) at discrete times $t_{i}$, with steps $\Delta t=t_{i}-t_{i-1}$, in
order to numerically integrate equations (\ref{xznorm})-(\ref{xznorm2}) with a
sufficient accuracy - provided that $\Delta t$ is small enough - owing to an
explicit integration scheme (\cite{VolokitinKrafft2020}).

\begin{figure*}
    \centering
    \includegraphics[width=0.6\textwidth]{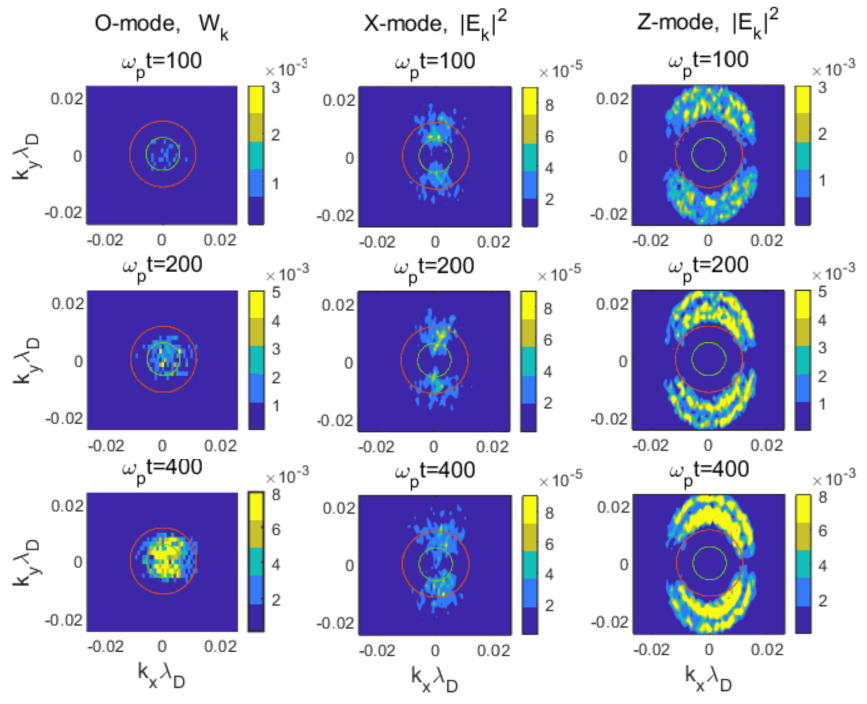}
    \caption{Energy spectra of the 
modes $\mathcal{O}$ ($W_{\mathbf{k}}=|\mathbf{E}_{\mathbf{k}}^{2}|+|\mathbf{B}_{\mathbf{k}}^{2}|$, left column), $\mathcal{X}$ ($|\mathbf{E}_{\mathbf{k}}^{2}$, middle column) and
$\mathcal{Z}$ ($|\mathbf{E}_{\mathbf{k}}^{2}|$, right column), in the map ($k_x\lambda_D,k_y\lambda_D$), at times $\omega_{p}t=100$
(upper row), $200$ (middle row) and $400$ (bottom row), for $\omega_{c}%
/\omega_{p}=0.15$ and $\Delta N=0.03.$ The circles represent the conditions
$3k^{2}\lambda_{D}^{2}=2\Delta N$ (red) and $3k^{2}\lambda_{D}^{2}=\Delta N$ (green).  Scales are linear. All
variables are normalized.}
    \label{fig8}
\end{figure*}

Fig. \ref{fig8} shows the evolution at different times of the spectral electromagnetic
energies of $\mathcal{O}$-, $\mathcal{X}$-
and $\mathcal{Z}$-modes, obtained using the same simulation performed in a
weakly magnetized plasma with $\omega_{c}/\omega_{p}=0.15$ and $\Delta N=0.03$.
Note that scales are linear so that small amplitudes appear as vanishing
values. The differences between the spectral energy distributions of modes, 
namely due to the terms $\Delta\omega_{\pm}$ in equations (\ref{xznorm}%
)-(\ref{xznorm2}) and $\Delta\omega_{\mathbf{k}}$ in (\ref{dbkdt}), have an essential impact on wave radiation. 
One observes that the main part of electromagnetic energy is carried by $\mathcal{Z}$-mode waves (\cite{Krafft2025}, at wavenumbers $3k_{\mathcal{Z}}^{2}\lambda_{D}^{2}\gtrsim2\Delta N$. On the contrary, $\mathcal{X}$-mode radiation is much weaker, whereas $\mathcal{O}$-mode emissions, at $3k_{\mathcal{O}}^{2}\lambda_{D}^{2}\lesssim\Delta N$, exhibit significantly higher energy levels without however reaching those of $\mathcal{Z}$-mode waves. These statements are confirmed below for a large set of plasma parameters (Figs. \ref{fig11}-\ref{fig14}).

\begin{figure}
    \centering
    \includegraphics[width=0.7\textwidth]{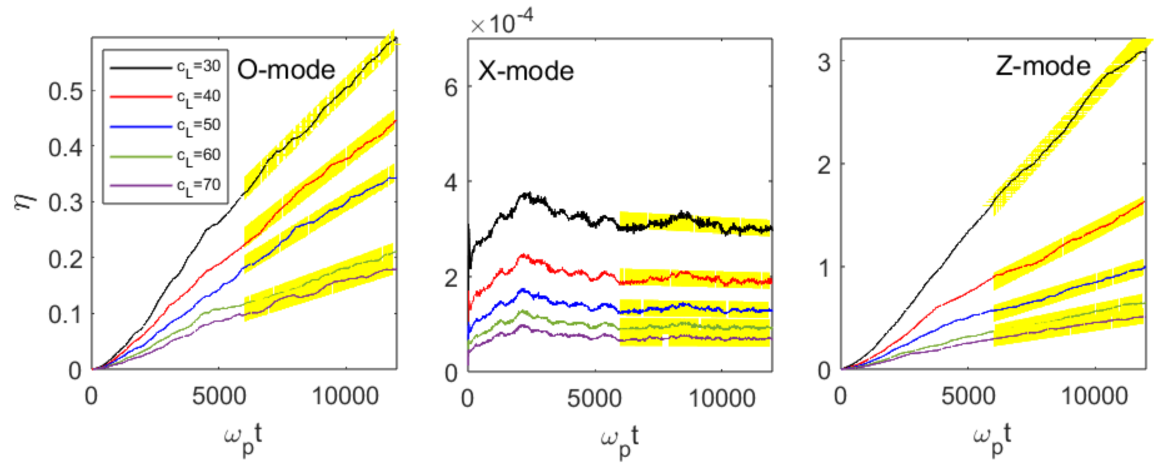}%Figures/OZX_1f_2_60327-FIGURE9.pdf}
    \caption{Time variations of the electromagnetic wave energy $\eta(t)$ in a
weakly magnetized plasma, for $\Delta N=0.03$, $\omega_{c}/\omega
_{p}=0.15$, and 5 values of $c_{L}$ (see legend in the left panel), for the
$\mathcal{O}$-mode (left), the $\mathcal{X}$-mode (middle), and the
$\mathcal{Z}$-mode (right). The superimposed thick yellow straight lines
highlight the slopes of the linear growths at large times, providing the time-independent
radiation rates $\dot{\eta}$.   All variables are normalized.}
    \label{fig9}
\end{figure}

\begin{figure}
    \centering
    \includegraphics[width=0.5\textwidth]{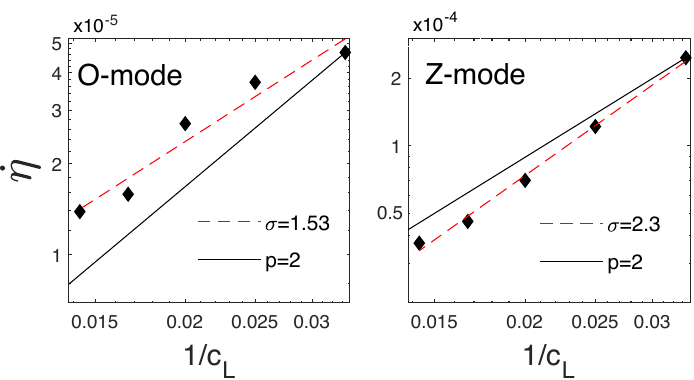}
    \caption{Variations of the electromagnetic radiation rate $\dot{\eta}$ as a function of $1/c_{L}$ (in logarithmic scales), for $\mathcal{O}$-mode (left)
and $\mathcal{Z}$-mode (right) waves, at plasma conditions of Fig. \ref{fig8}. The radiation rates calculated by the model equations are represented by black diamonds; linear interpolations are shown by
red dashed lines, indicating that $\sigma=1.53$ (left) and $\sigma
=2.3$ (right). The black lines show the scaling for the index $p=2$. All variables are normalized.}
    \label{fig10}
\end{figure}

\begin{figure}
    \centering
    \includegraphics[width=0.6\textwidth]{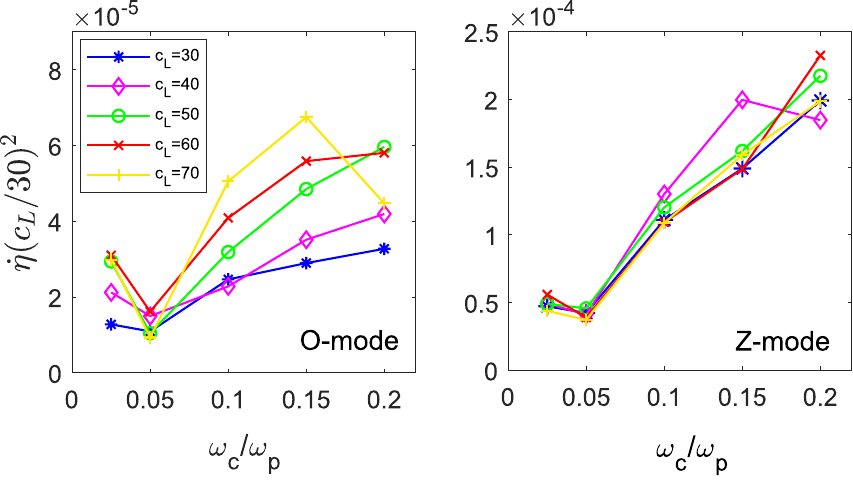}
    \caption{Variations with $\omega_{c}/\omega_{p}$ of the electromagnetic
radiation rate $\dot{\eta}(c_{L}/30)^{2}$ of $\mathcal{O}$-mode (left) and $\mathcal{Z}$-mode (right) waves, for $\Delta N=0.02$ and different
$c_{L}$ (see the legend in the left panel). All variables are normalized.}
    \label{fig11}
\end{figure}

\begin{figure}
    \centering
    \includegraphics[width=0.6\textwidth]{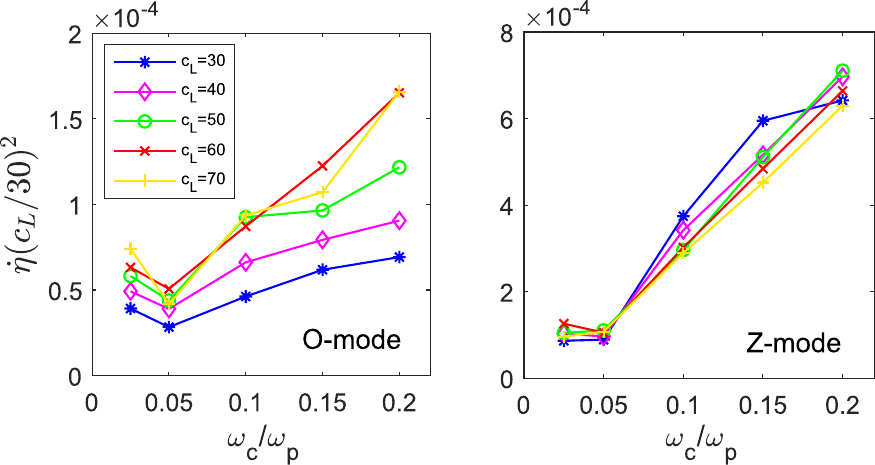}
    \caption{Variations with $\omega_{c}/\omega_{p}$ of the electromagnetic
radiation rate $\dot{\eta}(c_{L}/30)^{2}$ of $\mathcal{O}$-mode (left) and
 $\mathcal{Z}$-mode (right) waves, for $\Delta N=0.05$ and different 
$c_{L}$ (see the legend in the left panel). All variables are normalized.}
    \label{fig12}
\end{figure}

\begin{figure}
    \centering
    \includegraphics[width=0.6\textwidth]{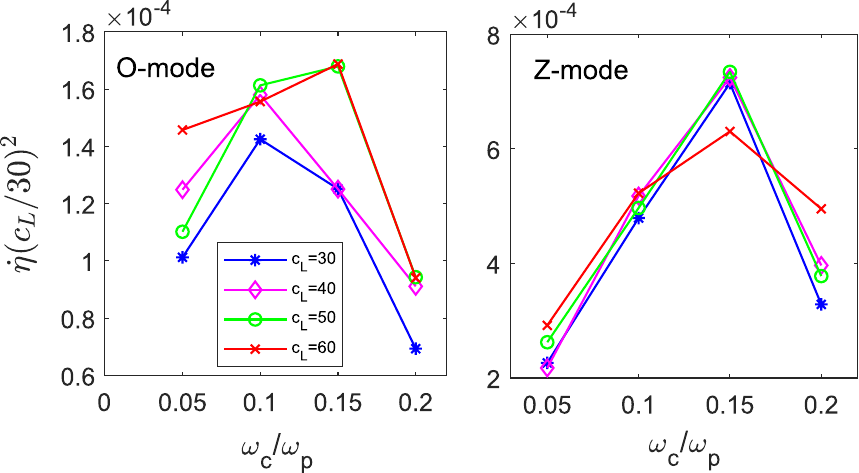}
    \caption{Variations with $\omega_{c}/\omega_{p}$ of the electromagnetic
radiation rate $\dot{\eta}(c_{L}/30)^{2}$ of  $\mathcal{O}$-mode (left) and
 $\mathcal{Z}$-mode (right) waves, for $\Delta N=0.06$ and different 
$c_{L}$ (see the legend in the left panel). All variables are normalized.}
    \label{fig13}
\end{figure}

\begin{figure}
    \centering
    \includegraphics[width=0.6\textwidth]{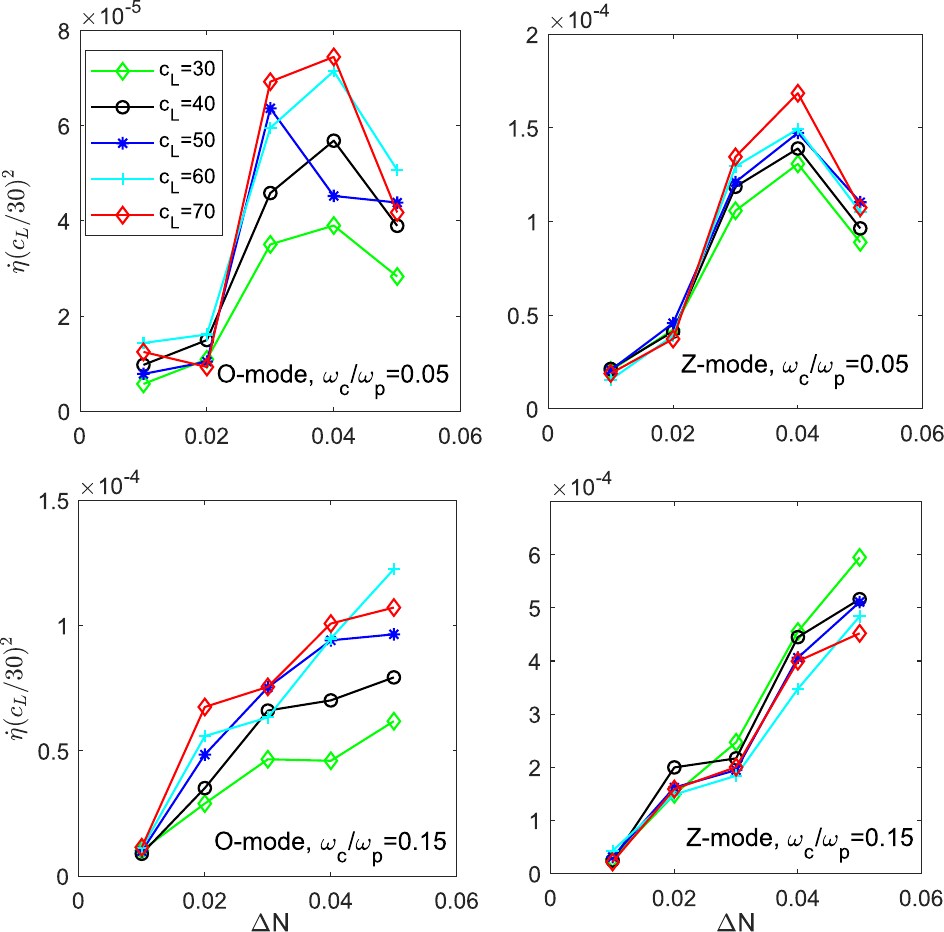}
    \caption{Variations with the average level of density fluctuations $\Delta N$
of the electromagnetic radiation rates $\dot{\eta}(c_{L}/30)^{2}$ of
$\mathcal{O}$-mode (left) and $\mathcal{Z}$-mode (right) waves, for $\omega
_{c}/\omega_{p}=0.05$ (upper row) and $\omega_{c}/\omega_{p}=0.15$ (bottom
row), for different $c_L$ (see the legend in the upper-left panel). All
variables are normalized.}
    \label{fig14}
\end{figure}

Fig. \ref{fig9} shows the time variations of the electromagnetic energy $\eta(t)$ carried by
each mode, confirming results provided by Fig. \ref{fig8}. For $\mathcal{O}$- and $\mathcal{Z}$-mode waves, the  radiation rates $\dot{\eta}$ decrease with increasing $c/v_T$; for $\mathcal{X}$-mode waves, they present weakly positive,  negative and vanishing values, showing that only negligibly small emissions are radiated. However, as studied in a previous work (\cite{Krafft2025}), the occurrence of $\mathcal{X}$-mode emission depends on the magnetization ratio $\omega_c/\omega_p$ and the average level of density fluctuations $\Delta N$. Indeed, for $\omega_c/\omega_p\lesssim\Delta N$, $\mathcal{X}$-mode waves can be radiated, whereas for $\omega_c/\omega_p\gtrsim\Delta N$, they do not emit significant energy. Therefore, we will compare below the radiation rates of $\mathcal{O}$- and $\mathcal{Z}$-mode waves only.
Fig. \ref{fig10} shows the variations of $\mathcal{O}$- and $\mathcal{Z}$-mode radiation rates $\dot{\eta}$  with $1/c_{L}$. Those reach typical values around
$\dot{\eta}_{\mathcal{O}}\simeq10^{-6}-10^{-5}$ and $\dot{\eta}_{\mathcal{Z}%
}\simeq10^{-5}-10^{-4}$, respectively, so that $\dot{\eta}_{\mathcal{Z}
}\sim10\dot{\eta}_{\mathcal{O}
}$ (\cite{Krafft2025}). For $\mathcal{O}$-mode waves, main
differences between the unmagnetized and the weakly magnetized plasma cases
concern the scaling indices of $1/c_L=v_T/c$, which show for the latter case larger
deviations from the index $p=2$  (compare with Figs. \ref{fig4} -\ref{fig6}). Note that 
analytic calculations performed in the framework of weak turbulence theory
predict, in 2D geometry, the scaling laws $\dot{\eta}_{\mathcal{Z}}%
\propto(v_{T}/c)^{2}$ and $\dot{\eta}_{\mathcal{O}}\propto(v_{T}/c)^{\sigma},$ with $1<\sigma<2$  (see section 3);  $\sigma$ is smaller when the plasma is magnetized than unmagnetized (compare with Figs. \ref{fig3}a, \ref{fig4}-\ref{fig6}).

Fig. \ref{fig11} shows that, for both $\mathcal{X}$- and $\mathcal{Z}$-mode waves, the dependence of $\dot{\eta}(c_L/30)^2$ on 
$\omega_{c}/\omega_{p}$ is close to linear if $\Delta N$ is
quite small, i.e. $\Delta N\lesssim0.02$; radiation rates grow with $\omega_{c}/\omega_{p}$. For  $\Delta N\gtrsim$ $\omega
_{c}/\omega_{p}$, radiation rates of $\mathcal{O}$- and $\mathcal{Z}$-mode
emissions reach larger values (Fig. \ref{fig12}, $\Delta N=0.05$), whereas for $\omega_{c}/\omega_{p}>\Delta N$,
$\mathcal{Z}$-mode waves exhibit a stronger linear dependence on $\omega_{c}%
/\omega_{p}$ than $\mathcal{O}$-mode ones and their radiation rates are significantly higher. As $\Delta N$ increases to $\Delta N=0.06$ (Fig. \ref{fig13}), the linear dependence on
$\omega_{c}/\omega_{p}$  breaks and radiation rates of both modes
present the same behavior with a maximum around $\omega_{c}/\omega_{p}%
\simeq0.15$.

Fig. \ref{fig14} shows the variations with $\Delta N$ of the radiation rates
$\dot{\eta}(c_{L}/30)^{2}$ of $\mathcal{O}$- and $\mathcal{Z}$- modes, for
$\omega_{c}/\omega_{p}=0.05$ and $\omega_{c}/\omega_{p}=0.15$, as well as for
different values of $c_{L}.$ At $\Delta N\lesssim\omega_{c}/\omega_{p}$ (upper
row), radiation rates exhibit a maximum around $\Delta
N\simeq0.04$ (upper row), whereas for higher $\omega_{c}/\omega_{p}=0.15$
($\Delta N\ll\omega_{c}/\omega_{p}$), they grow quasi-linearly with
$\Delta N$ (bottom row)$.$ We again observe that radiation rates of
$\mathcal{Z}$-mode waves are always larger than those of $\mathcal{O}%
$-mode ones; both increase with $\omega_{c}/\omega_{p}$, in agreement with Figs.  \ref{fig11}- \ref{fig12}.

\section{Analytical determination of electromagnetic radiation rates}
Let us determine analytically the rates at which, in a weakly magnetized
plasma with random density fluctuations, the energy of electrostatic
upper-hybrid waves is transformed, at constant frequency, into
electromagnetic radiation at $\omega _{p}$, in the $\mathcal{O}$-, 
$\mathcal{X}$- and $\mathcal{Z}$-modes. The same approximations as done
above for the determination of equations governing electromagnetic radiation are used, regarding the
properties of density fluctuations and upper-hybrid wave turbulence (in particular, its intensity). In addition, further assumptions are included, the validity of
which is based on the results presented above.

Note that in this section we use non normalized variables and cgs units. As
the plasma magnetization is weak, we neglect terms of the second order in $%
\omega _{c}.$ Calculations are performed in the framework of 3D geometry.
The ambient magnetic field is directed along the unitary vector $\mathbf{h}$.
Let us start from the general equations (\ref{dBkdt})-(\ref{G}), which provide that 

\begin{equation}
\left( i\frac{\partial }{\partial t}-\Delta \omega _{\mathbf{k}}\right) B_{%
\mathbf{k}}=-\frac{c}{2}\frac{\omega _{p}^{2}}{\omega _{p}^{2}-\omega
_{c}^{2}}\mathbf{a}_{\mathbf{k}}^{\ast }\cdot \mathbf{G_{\mathbf{k}}} 
=-\frac{%
c}{2}\frac{\omega _{p}^{2}}{\omega _{p}^{2}-\omega _{c}^{2}}\sum_{\mathbf{k}=%
\mathbf{k}_{1}\mathbf{+k}_{2}}\left( \mathbf{a}_{\mathbf{k}}^{\ast }\cdot 
\mathbf{\beta }_{\mathbf{kk}_{2}}\right) \rho _{\mathbf{k}_{1}}\varphi _{%
\mathbf{k}_{2}},  \label{Eq3}
\end{equation}
where $\Delta \omega _{\mathbf{k}}=\omega _{\mathbf{k}}-\omega _{p}$; $%
\varphi _{\mathbf{k}},$ $\rho _{\mathbf{k}}$ and $\mathbf{G}_{\mathbf{k}}$
are the Fourier components of the potential envelope $\tilde{\varphi},$ the
normalized density perturbation $\rho =\delta n/n_{0}$ and the vector $%
\mathbf{G}$ (\ref{G}) proportional to $\nabla \times \delta \mathbf{\tilde{j}}$; $\mathbf{G_{\mathbf{k}}}$ is expressed though the vector $\mathbf{\beta }_{\mathbf{kk}_{2}}$ (see below) that
contains $\mathbf{k}$ and $\mathbf{k}_{2}$ and depends on the
electromagnetic mode considered.  The full solution of (\ref{Eq3}) is given at time $t$ by 

\begin{equation}
B_{\mathbf{k}}(t)=B_{\mathbf{k}}(0)e^{-i\Delta \omega _{\mathbf{k}}t}
+\frac{c%
}{2}\frac{\omega _{p}^{2}}{\omega _{p}^{2}-\omega _{c}^{2}}\sum_{\mathbf{k=k}%
_{1}+\mathbf{k}_{2}}\left( \mathbf{a}_{\mathbf{k}}^{\ast }\cdot \mathbf{%
\beta }_{\mathbf{kk}_{2}}\right) \int_{0}^{t}\rho _{\mathbf{k}_{1}}\left(
t^{\prime }\right) \varphi _{\mathbf{k}_{2}}(t^{\prime })e^{i\Delta \omega _{%
\mathbf{k}}\left( t^{\prime }-t\right) }dt^{\prime }.  \label{Bkformal}
\end{equation}

Considering the evolution of $\mathbf{B}%
_{\mathbf{k}}\left( t\right) $ at large times (see section 2), we can
neglect the small initial values $B_{\mathbf{k}}(0)$; then, squaring (\ref%
{Bkformal}), we get

\begin{equation}
\left\vert B_{\mathbf{k}}(t)\right\vert ^{2}\simeq \left( \frac{c}{2}\frac{%
\omega _{p}^{2}}{\omega _{p}^{2}-\omega _{c}^{2}}\right) ^{2}
\sum_{\mathbf{%
k=k}_{1}+\mathbf{k}_{2}}\sum_{\mathbf{k=k}_{3}+\mathbf{k}_{4}}\left( \mathbf{%
a}_{\mathbf{k}}^{\ast }\cdot \mathbf{\beta }_{\mathbf{kk}_{2}}\right) \left( 
\mathbf{a}_{\mathbf{k}}\cdot \mathbf{\beta }_{\mathbf{kk}%
_{4}}^{\ast }\right)
\int_{0}^{t}dt^{\prime }\int_{0}^{t}dt^{\prime \prime }\rho _{\mathbf{%
k}_{1}}\left( t^{\prime }\right) \rho _{\mathbf{k}_{3}}^{\ast }\left(
t^{\prime \prime }\right) \varphi _{\mathbf{k}_{2}}(t^{\prime })\varphi _{%
\mathbf{k}_{4}}^{\ast }(t^{\prime \prime })e^{i\Delta \omega _{\mathbf{k}%
}\left( t^{\prime }-t^{\prime \prime }\right) }.  \label{Bk22}
\end{equation}

The quantity $\left\vert \mathbf{B}_{\mathbf{k}}\left( t\right) \right\vert
^{2}$ naturally experiences statistical fluctuations. We are interested in
its growth with time on average. In our case, when density fluctuations are
quasi-static (i.e., the dependence of density fluctuations' amplitudes on
time can be neglected) and completely random, we can assume that $\left\langle\rho_{\mathbf{k}_{1}}\left( t^{\prime }\right)\rho_{\mathbf{k}_{3}}^{\ast }\left( t^{\prime \prime }\right)\right\rangle \simeq\delta _{\mathbf{k}_{1}\mathbf{k}_{3}}\left\vert\rho_{\mathbf{k}_{1}}\right\vert ^{2}$, 
where the brackets denote
statistical averaging, supposed to be consistent with time
averaging. Further
calculations are only possible with such additional hypotheses, as done in the framework of weak turbulence theory. After averaging over
the ensemble of density fluctuations, we obtain

\begin{equation}
\left\langle \left\vert B_{\mathbf{k}}(t)\right\vert ^{2}\right\rangle
\simeq \left( \frac{c}{2}\frac{\omega _{p}^{2}}{\omega _{p}^{2}-\omega
_{c}^{2}}\right) ^{2}\sum_{\mathbf{k=k}_{1}+\mathbf{k}_{2}}\left\vert 
\mathbf{a}_{\mathbf{k}}^{\ast }\cdot \mathbf{\beta }_{\mathbf{kk}%
_{2}}\right\vert ^{2}\left\vert \rho _{\mathbf{k}_{1}}\left( t\right)
\right\vert ^{2}
\left\langle \int_{0}^{t}\left( \int_{0}^{t}\varphi _{%
\mathbf{k}_{2}}(t^{\prime })\varphi _{\mathbf{k}_{2}}^{\ast }(t^{\prime
\prime })\exp \left( i\Delta \omega _{\mathbf{k}}\left( t^{\prime
}-t^{\prime \prime }\right) \right) dt^{\prime \prime }\right) dt^{\prime
}\right\rangle ,
\end{equation}

where we took the term $\left\vert \rho_{\mathbf{k}_{1}}\left( t\right)\right\vert ^{2}$ from the time integrals due to its slower variation. To go further, we assume that the correlations between
amplitudes of upper-hybrid waves decay exponentially as
\begin{equation}
\left\langle \varphi _{\mathbf{k}_{2}}(t^{\prime })\varphi _{\mathbf{k}%
_{2}}^{\ast }(t^{\prime \prime })\right\rangle =\left\vert \varphi _{\mathbf{%
k}_{2}}\right\vert ^{2}\exp \left( -\nu _{\mathbf{k}_{2}}\left\vert
t^{\prime }-t^{\prime \prime }\right\vert -i\delta \omega _{\mathbf{k}%
_{2}}\left( t^{\prime }-t^{\prime \prime }\right) \right), \label{phiphi}
\end{equation}
where $\delta \omega _{\mathbf{k}_{2}}=\omega _{\mathbf{k}%
_{2}}-\omega _{p}$ ($\omega _{\mathbf{k}_{2}}$ is the frequency of
upper-hybrid waves); the frequency $\nu _{\mathbf{k}_{2}}>0$ is determined
by the interactions of these waves with random density fluctuations.
Then we get 
\begin{multline}
\left\langle \left\vert B_{\mathbf{k}}(t)\right\vert ^{2}\right\rangle
\simeq \left( \frac{c}{2}\frac{\omega _{p}^{2}}{\omega _{p}^{2}-\omega
_{c}^{2}}\right) ^{2}\sum_{\mathbf{k=k}_{1}+\mathbf{k}_{2}}\left\vert 
\mathbf{a}_{\mathbf{k}}^{\ast }\cdot \mathbf{\beta }_{\mathbf{kk}%
_{2}}\right\vert ^{2}\left\vert \rho _{\mathbf{k}-\mathbf{k}_{2}}\left(
t\right) \right\vert ^{2}\left\vert \varphi _{\mathbf{k}_{2}}(t)\right\vert
^{2}\times\\ 
\times\int_{0}^{t} \int_{0}^{t}\exp \left( i\left( \Delta
\omega _{\mathbf{k}}-\delta \omega _{\mathbf{k}_{2}}\right) \left( t^{\prime
}-t^{\prime \prime }\right) -\nu _{\mathbf{k}_{2}}\left\vert t^{\prime
}-t^{\prime \prime }\right\vert\right) dt^{\prime \prime}dt^{\prime}, \label{Itt}
\end{multline}

where the double integral in the rhs term tends to 
${2\nu _{\mathbf{k}_{2}}t}[{\left( \Delta
\omega _{\mathbf{k}}-\delta \omega _{\mathbf{k}_{2}}\right) ^{2}+\nu _{%
\mathbf{k}_{2}}^{2}}]^{-1}$ at large times $t$, which in turn can be approximated by the Dirac function $2\pi
t\delta \left( \Delta \omega _{k}-\delta \omega _{\mathbf{k}_{2}}\right) $\
if $\nu _{\mathbf{k}_{2}}$ is not too large. Then, the radiation rate of magnetic
energy is given by

\begin{equation}
\frac{d}{dt}\left\langle \left\vert B_{\mathbf{k}}(t)\right\vert
^{2}\right\rangle \simeq 2\pi \left( \frac{c}{2}\frac{\omega _{p}^{2}}{%
\omega _{p}^{2}-\omega _{c}^{2}}\right) ^{2} 
\sum_{\mathbf{k=k}_{1}+\mathbf{k}%
_{2}}\left\vert \mathbf{a}_{\mathbf{k}}^{\ast }\cdot \mathbf{\beta }_{%
\mathbf{kk}_{2}}\right\vert ^{2}\left\vert \rho _{\mathbf{k}-\mathbf{k}%
_{2}}\left( t\right) \right\vert ^{2}\left\vert \varphi _{\mathbf{k}%
_{2}}(t)\right\vert ^{2}\delta \left( \omega _{\mathbf{k}}^{t}-\omega _{%
\mathbf{k}_{2}}\right) %, \label{dB2}
\end{equation}

where $\Delta \omega _{\mathbf{k}}-\delta \omega _{\mathbf{k}_{2}}=\omega _{%
\mathbf{k}}^{t}-\omega _{\mathbf{k}_{2}}$, $\omega _{\mathbf{k}}^{t}$
designing the frequency of the transverse electromagnetic waves radiated at $%
\omega _{p}$. Then,  summing the entire electromagnetic wave
spectrum, taking into account that $\left\vert \mathbf{k}\right\vert \ll
\left\vert \mathbf{k}_{1}\right\vert ,\left\vert \mathbf{k}_{2}\right\vert $
, i.e. $\mathbf{k}_{2}\simeq -\mathbf{k}_{1},$ and assuming that spectra of
waves and density fluctuations are sufficiently smooth, we can sum over $%
\mathbf{k}_{2}$\ and $\mathbf{k}$ independently and get the radiation rate
at $\omega _{p}$ 
\begin{equation}
\frac{d}{dt}\sum_{\mathbf{k}}\left\langle \left\vert B_{\mathbf{k}%
}(t)\right\vert ^{2}\right\rangle \simeq \frac{\pi c^{2}}{2}\left( \frac{%
\omega _{p}^{2}}{\omega _{p}^{2}-\omega _{c}^{2}}\right) ^{2}\sum_{\mathbf{k}%
_{2}}{\left\vert \varphi_{\mathbf{k}_{2}}(t)\right\vert ^{2}}
\left( \sum_{\mathbf{k}}\left\vert \rho _{\mathbf{k}-\mathbf{k}_{2}}\left(
t\right) \right\vert ^{2}\left\vert \mathbf{a}_{\mathbf{k}}^{\ast }\cdot 
\mathbf{\beta }_{\mathbf{kk}_{2}}\right\vert ^{2}\delta \left( \omega _{%
\mathbf{k}}^{t}-\omega _{\mathbf{k}_{2}}\right) \right)  \label{Bk1}
\end{equation}

Then, using that $V_{s}^{-1}\sum_{\mathbf{k}}f(\mathbf{k})=\int_{V_{s}} f(%
\mathbf{k})d^{s}\mathbf{k}/(2\pi )^{s}$, where $V_{s}$ is the volume
of the system of dimension $s$, we obtain
\begin{equation}
\frac{d}{dt}\int_{V_{s}}\left\langle \left\vert B_{\mathbf{k}}(t)\right\vert
^{2}\right\rangle d^{s}\mathbf{k}\simeq \frac{V_{s}}{(2\pi )^{s}}\frac{\pi
c^{2}}{2}\left( \frac{\omega _{p}^{2}}{\omega _{p}^{2}-\omega _{c}^{2}}%
\right) ^{2}
\int_{V_{s}}\left\vert E_{\mathbf{k}_{2}}(t)\right\vert
^{2}I\left( \mathbf{k}_{2}\right) d^{s}\mathbf{k}_{2},
  \label{rate_general}
\end{equation}
where we replaced the potentials $\varphi _{\mathbf{k}_{2}}$ with the electric fields $E_{\mathbf{k}_{2}}$, and 
\begin{equation}
I\left( \mathbf{k}_{2}\right) =\int_{V_{s}}\left\vert 
\mathbf{a}_{\mathbf{k}}^{\ast }\cdot \mathbf{\beta }_{\mathbf{kk}%
_{2}}^{\prime}\right\vert ^{2}\left\vert \rho _{\mathbf{k}-\mathbf{k}_{2}}\left(t\right) \right\vert ^{2}\delta \left( \omega _{\mathbf{k}}^{t}-\omega _{%
\mathbf{k}_{2}}\right){k}^{2} d^{s}\mathbf{k}
,  \label{Ik2}
\end{equation}
where we defined $\left\vert 
\mathbf{a}_{\mathbf{k}}^{\ast }\cdot \mathbf{\beta }_{\mathbf{kk}%
_{2}}^{\prime }\right\vert ^{2}=\left\vert \mathbf{a}_{\mathbf{k}}^{\ast
}\cdot \mathbf{\beta }_{\mathbf{kk}_{2}}\right\vert ^{2}/k^{2}k_{2}^{2}.$
These general and rather simple equations  (\ref{rate_general})-(\ref{Ik2})  allow to calculate the radiation rates of any electromagnetic mode, if wave and density turbulence spectra $\left\vert E_{\mathbf{k}_{2}}(t)\right\vert
^{2}$ and $\left\vert \rho _{\mathbf{k}-\mathbf{k}_{2}}\left(t\right) \right\vert ^{2}$ are known, together with the dispersion and the polarization properties of electrostatic and electromagnetic waves.
Below we apply them to the case of the radiation of electromagnetic ordinary and extraordinary modes by upper-hybrid wave turbulence.

\subsection{Radiation of electromagnetic ordinary mode waves}
Let us apply equations (\ref{rate_general})-(\ref{Ik2}) to the calculation of the radiation rate of electromagnetic ordinary mode waves.
Their magnetic energy is mostly carried by the magnetic component 
perpendicular to $\mathbf{B}_{0}$. Moreover, it follows from $\nabla \cdot 
\mathbf{B}=0$ that $\mathbf{k\cdot B}_{\mathbf{k}}=0$. Thus,  the corresponding
polarization vector can be approximated by  $\mathbf{a}_{\mathbf{k}}\simeq{\mathbf{k\times h}}/{\left\vert \mathbf{k\times h}\right\vert }={\mathbf{k\times h}}/{k_{\perp }}$. Note that this expression is not accurate enough for parallel and quasi-parallel
wave propagation. However, such waves contribute insignificantly to
electromagnetic radiation, so that we can neglect the inaccuracy of the polarization vector. 
Note that in a magnetized plasma, the $\mathcal{O}$-mode spectrum is no longer a circle (as for $\omega_c=0)$, 
but presents a significant  anisotropy (see  Fig. \ref{fig15}  of Appendix $B$), which has
to be taken into account when calculating $I\left( \mathbf{k}_{2}\right) $. 
Equations (\ref{G})-(\ref{gzk}) obtained in  section 2 provide that 
\begin{equation}
\mathbf{\beta }_{\mathbf{kk}_{2}}\simeq i(\mathbf{k}\times \mathbf{k}_{2}+i%
\frac{\omega _{c}}{\omega }\mathbf{k}\times \left( \mathbf{h}\times \mathbf{k%
}_{2}\right)) ,  \label{beta}
\end{equation}
so that we can calculate
\begin{equation}
\left\vert \mathbf{a}_{\mathbf{k}}^{\ast }\cdot \mathbf{\beta^{\prime} }_{\mathbf{kk}%
_{2}}\right\vert ^{2}\simeq\frac{k_{\parallel }^{2}}{k_{\perp }^{2}k^{2}k_{2 }^{2}}\left( \left(
k^{2}\frac{k_{2\parallel }}{k_{\parallel }}-\left( \mathbf{k}\cdot \mathbf{k}%
_{2}\right) \right) ^{2}\mathbf{+}\frac{\omega _{c}^{2}}{\omega ^{2}}\left( 
\mathbf{h}\cdot \left( \mathbf{k}_{2}\times \mathbf{k}\right) \right)
^{2}\right)   \label{ak-betak}
\end{equation}
Using spherical coordinates ($k,\theta ,\psi $)  and neglecting the second order terms in $\ \omega _{c}^{2},$ we obtain 
\begin{equation}
\left\vert \mathbf{a}_{\mathbf{k}}^{\ast }\cdot \mathbf{\beta }_{\mathbf{kk}%
_{2}}^{\prime }\right\vert ^{2}\simeq\frac{\left( \cos \theta _{2}-\cos \theta
\cos \alpha \right) ^{2}}{\sin ^{2}\theta}, \label{akprime}
\end{equation}
with  $\cos \alpha ={\mathbf{k}\cdot \mathbf{k}_{2}}/{kk_{2}}$. Expressing equation (\ref{akprime}) using angles $\theta ,\psi $, and $\theta _{2},\psi _{2}$,
and integrating on $\psi $, we get    
\begin{equation}
\int_{0}^{2\pi }\left\vert \mathbf{a}_{\mathbf{k}}^{\ast }\cdot \mathbf{%
\beta }_{\mathbf{kk}_{2}}^{\prime }\right\vert ^{2}d\psi \simeq \left( 2\cos
^{2}\theta _{2}\sin ^{2}\theta +\cos ^{2}\theta \sin ^{2}\theta _{2}\right) 
\label{Int-psi},
\end{equation}
so that 
\begin{equation}
I\left( \mathbf{k}_{2}\right) =\frac{1 }{8\pi ^{2}}%
\int_{0}^{\pi }\left( 2\cos ^{2}\theta _{2}\sin ^{2}\theta +\cos ^{2}\theta
\sin ^{2}\theta _{2}\right) \left( \int_{0}^{\infty }\delta \left( \omega _{%
\mathbf{k}}^{t}-\omega _{\mathbf{k}_{2}}\right) k^{4}dk\right) \sin \theta
d\theta.   \label{Ik2-2}
\end{equation}
The dispersion relation of $\mathcal{O}$-mode waves in a weakly magnetized
plasma can be approximated by  
\begin{equation}
\omega \simeq \omega _{p}+\frac{k^{2}c^{2}}{2\omega _{p}}\sin ^{2}\theta \label{disp1t}
\end{equation}
for $k^{2}c^{2}\sin ^{2}\theta \leq \omega _{p}\omega _{c}$ ($\sin^{2}\theta \neq 0$),  and by
\begin{equation}
\omega \simeq \omega _{p}+\frac{\omega _{c}^{2}-\omega _{p}\omega _{c}\cos
^{2}\theta }{2\omega _{p}}+\frac{k^{2}c^{2}}{2\omega _{p}}   \label{disp2t}
\end{equation}
for larger wavenumbers. Such splitting is required due to a singularity in the $\mathcal{O}$-mode dispersion (see Appendix $B$ for more details, and Fig.\ref{fig16}).   Then,  defining $k_{0}=\sqrt{3}k_{2}v_{T}/c$ ($\mathcal{O}$-mode wavenumber in an unmagnetized plasma), we can write for the first $k$-range (\ref{disp1t}) that  
\begin{equation}
\delta \left( \omega _{\mathbf{k}}^{t}-\omega _{\mathbf{k}_{2}}\right)
=\delta \left( \frac{c^{2}}{2\omega _{p}}\left( k^{2}\sin ^{2}\theta
-k_{0}^{2}\right) \right) ,  \label{direc1}
\end{equation}
and, for the second $k$-range (\ref{disp2t}), that 
\begin{equation}
\delta \left( \omega _{\mathbf{k}}^{t}-\omega _{\mathbf{k}_{2}}\right)
=\delta \left( \frac{c^{2}}{2\omega _{p}}\left( k^{2}-K^2(\theta)
\right) \right) ,  \label{dirac2}
\end{equation}
where $ K^2(\theta)=k_{0}^{2}+ (\omega _{p}\omega _{c}\cos ^{2}\theta -\omega
_{c}^{2})/c^{2}$. For the first case with dispersion (\ref{disp1t}), we replace the spherical coordinates $(k, \theta) $ by the cylindrical ones $(k_{\perp}, k_{\parallel}) $,  integrate on $k_{\perp }$ and finally obtain (\ref{Ik2-2}) in the form 
\begin{equation}
I\left( \mathbf{k}_{2}\right)=\frac{\pi \omega _{p}}{c^{2}}J(\theta_2,k_2,k_0) \\ 
\simeq\frac{\pi \omega _{p}}{c^{2}}%
\int_{-\infty }^{\infty }\left\vert {\rho_{\mathbf{k}(k_{0,}k_{\parallel})-\mathbf{k}%
_{2}}\left( t\right) }\right\vert ^{2}\left(
\frac{6k_{2}^{2}v_{T}^{2}}{c^{2}}\cos ^{2}\theta _{2}+k_{\parallel }^{2}\sin ^{2}\theta _{2}\right)
dk_{\parallel }. \label{Ik2-0}
\end{equation}
We get then from equation (\ref{rate_general})  the first contribution to the radiation rate 
\begin{equation}
\dot{\mu}_{{\mathcal{O},1}}=\frac{d}{\omega _{p}dt}\int_{V}\left\langle\left\vert B_{\mathbf{k}%
}(t)\right\vert ^{2}\right\rangle \frac{d^{3}\mathbf{k}}{(2\pi
)^{3}}  \\
\simeq \frac{V}{16\pi }\left( \frac{\omega _{p}^{2}}{\omega _{p}^{2}-\omega
_{c}^{2}}\right) ^{2} \int_{V}\frac{d^{3}\mathbf{k}_2}{(2\pi
)^{3}} \left\vert E_{\mathbf{k}
_{2}}(t)\right\vert ^{2}J(\theta_2,k_2,k_0). \label{Ik2-1}
\end{equation}
The integral in the rhs of equation (\ref{Ik2-1}) is not singular, due to the
presence of the density spectrum $\left\vert {%
\rho_{\mathbf{k}(k_{0,}k_{\parallel})-\mathbf{k}_{2}}\left(t\right) }\right\vert ^{2}$ which is vanishing outside the plasma source, and thus when $k_{\parallel}$ tends to infinity. An analytical integration  can be easily  performed with a Gaussian density spectrum, for example.
Note that the density spectrum cannot be taken out of the integrand, resulting from the fact that  $\mathbf{k}$ cannot be neglected compared to $\mathbf{k_2}$ in the considered $k$-range, due to some specific features of the $\mathcal{O}$-mode wave dispersion relation (see Appendix $B$). 

The second contribution to the radiation rate, corresponding to the dispersion relation (\ref{disp2t}), provides 
\begin{equation}
I\left( \mathbf{k}_{2}\right) =\frac{\pi\omega _{p}}{c^{5}}\left(
\omega _{p}\omega _{c}\right) ^{3/2}\left\vert {\rho_{-\mathbf{k}%
_{2}}\left( t\right) }\right\vert ^{2} J(\theta _{2},k_{2}),
\end{equation}
and
\begin{equation}
J(\theta _{2},k_{2})=\left( 1-3\cos ^{2}\theta
_{2}\right) \int_{-1}^{1}\left( \cos \theta ^{2}+b\right) (\cos
\theta ^{2}+a)^{3/2}d(\cos \theta ), \label{J2}
\end{equation}
with $a={3k_{2}^{2}\lambda _{D}^{2}}\omega _{p}/\omega _{c}-{\omega
_{c}}/{\omega _{p}}$ and  $b={2\cos ^{2}\theta _{2}}/(1-3\cos ^{2}\theta _{2})$. The integration of equation (\ref{J2}) can be performed analytically (\cite{GradshteynRyzhik2007}), reducing $J(\theta_2,k_2)$ to simple forms in the limiting cases $a\ll1$ and $a\gg 1$. Finally, we get the second part of the $\mathcal{O}$-mode radiation rate in the form
\begin{equation}
\dot{\mu}_{{\mathcal{O},2}}\simeq \frac{V}{\lambda
_{D}^{3}}\frac{1}{32\pi }\left( \frac{\omega _{c}}{\omega _{p}}\right)
^{3/2}\left( \frac{v_{T}}{c}\right) ^{3}
\int_{V} \frac{d^{3}\mathbf{k}_2}{(2\pi)^{3}}\left\vert E_{\mathbf{k}%
_{2}}(t)\right\vert ^{2}\left\vert {\rho_{-\mathbf{k}_{2}}\left(
t\right) }\right\vert ^{2}J(\theta _{2},k_{2}). \label{Ik2-3}
\end{equation}
Generally, the total  $\mathcal{O}$-mode radiation rate $\dot{\mu}_{{\mathcal{O}}} $ includes the contributions of the two $k$-regions (\ref{disp1t})-(\ref{disp2t}), i.e.  the sum of equations (\ref{Ik2-1}) and (\ref{Ik2-2}). The resulting expression $\dot{\mu}_{{\mathcal{O}}}=\dot{\mu}_{{\mathcal{O}},1} + \dot{\mu}_{{\mathcal{O}},2}$  shows that the total radiation rate does not scale as $(v_T/c)^3$; indeed, it exhibits two terms, containing $(v_T/c)^3$ (\ref{Ik2-2}) and $(v_T/c)^2$ (\ref{Ik2-1}), respectively; the actual scaling index is then between 2 and 3   in 3D geometry.  This explains why, in 2D geometry, we observe $\mathcal{O}$-mode radiation rates in a magnetized plasma with scaling indices between 1 and 2 (see Fig. \ref{fig10}). 
When the plasma is not magnetized, the second contribution  $\dot{\mu}_{{\mathcal{O}},2} $  (\ref{Ik2-3}) vanishes; in this case, the scaling index of $v_T/c$ is $\sigma\sim2$ in 2D and 3D geometry, in agreement with Figs. \ref{fig3}-\ref{fig6}.

\subsection{Radiation rate of electromagnetic extraordinary mode waves}
Let us now determine the radiation rates of extraordinary modes. A  general calculation was performed by the authors in 3D geometry in a previous work (\cite{Krafft2025}). However, let us start here from equations (\ref{xznorm})-(\ref{xznorm2}),
obtained in our model to describe the radiation of $\mathcal{X}$- and $\mathcal{Z}$-modes in 2D geometry, which have a form close to (\ref{Eq3}), including the electric instead of the wave magnetic field 
\begin{equation}
\left( i\frac{\partial }{\partial t}-\Delta \omega _{\mathbf{k}}^{\pm
}\right) E_{\mathbf{k}}^{\pm }\left( t\right) =\frac{i\omega _{p}^{2}}{%
2\left( \omega _{c}\mp \omega \right) }\left( \frac{\delta n}{n_{0}}\frac{%
\partial \tilde{\varphi}}{\partial y}\right) _{\mathbf{k}}.
\end{equation}
Note that variables are not normalized, that the signs "$+$" and "$-$"  correspond to $\mathcal{X}$- and $\mathcal{Z}$-modes, respectively, and that 
calculations performed with the electric field allow to avoid the use of polarization vectors (see (\cite{Krafft2025}) for magnetic energy radiation rates). It is more suitable technically to use electric fields here, as two symmetric equations are obtained above for $E_{\mathbf{k}}^{\pm }$ (\ref{envelopes}).
Therefore, using (\ref{Eq3})-(\ref{Ik2})  and replacing $\left\vert \mathbf{a}_\mathbf{k}^*\cdot\beta
_{\mathbf{k}_{1}\mathbf{k}_{2}}\right\vert ^{2}$ with $k_{2\perp }^{2}$, we get in 2D cylindrical coordinates $(k,\theta)$ that

\begin{equation}
\frac{d}{dt}\int_{V} \frac{d^{2}\mathbf{k}}{(2\pi )^{2}}\left\langle \left\vert
E_{\mathbf{k}}^{\pm }(t)\right\vert ^{2}\right\rangle \simeq \frac{V}{%
2(2\pi )^{2}}\left( \frac{\omega _{p}^{2}}{2\left( \omega _{c}\mp \omega
\right) }\right) ^{2}
\int_{V} \frac{d^{2}\mathbf{k}_{2}}{(2\pi )^{2}}\left\vert 
{\rho_{-\mathbf{k}_{2}}\left( t\right) }\right\vert ^{2}%
{\left\vert E_{\mathbf{k}_{2}}(t)\right\vert ^{2}}%
\sin ^{2}\theta _{2}
\int_{\Omega }\int_{0}^{\infty }\delta \left(
\omega _{\mathbf{k}}^{t}-\omega _{\mathbf{k}_{2}}\right)
k^{s-2}dk^{2}d^{s-1}\Omega,
\end{equation}

where $d\Omega=\sin{\theta}d\theta$; $\Omega$ is the angular domain of integration.  
Using the dispersion of $\mathcal{X}$- and $\mathcal{Z}$-modes near their
cutoff frequencies (see Appendix B and (\ref{disp})) and defining  $%
g_{\theta }=1+\cos ^{2}\theta $, we get 
\begin{equation}
\delta \left( \omega _{\mathbf{k}}^{t}-\omega _{\mathbf{k}_{2}}\right)
=\delta \left( \frac{c^{2}g_{\theta }}{4\omega _{p}}%
\left( k^{2}-\frac{k_{\pm }^{2}}{g_{\theta } }\right) \right) ,
\;\;\;\;
k_{\pm }^{2}\lambda _{D}^{2}=\frac{2v_{T}^{2}}{c^{2}}\left(
3k_{2}^{2}\lambda _{D}^{2}\mp \frac{\omega _{c}}{\omega _{p}}\right) .
\label{kpm}
\end{equation}
Equation (\ref{kpm}) requires that $k_{\pm }^{2}\lambda _{D}^{2}>0$ for the
radiation rate to be positive. This condition is always satisfied for the $%
\mathcal{Z}$-mode but, for the $\mathcal{X}$-mode, it is only fulfilled for
plasmas with $\omega _{c}/\omega _{p}<3k_{2}^{2}\lambda _{D}^{2}$ (\cite{Krafft2025}). Then, the
radiation rates $\dot{\eta}^{\pm }$ of $\mathcal{X}$- and $\mathcal{Z}$
-modes can be written in 2D geometry as
\begin{equation}   
\dot{\eta}^{\pm }_{2D}=\frac{d}{\omega _{p}dt}\int_V \frac{d^{2}\mathbf{k}}{(2\pi
)^{2}}\left\langle \left\vert E_{\mathbf{k}}^{\pm }(t)\right\vert
^{2}\right\rangle \\ 
\simeq \frac{1}{8\pi ^{2}}\frac{V}{\lambda _{D}^{2}}%
{\left(\frac{v_T}{c}\right)}^{2}\frac{\mathcal{J}}{\left( 1\mp \omega _{c}/\omega_{p}\right)^{2}}\int_V \frac{d^{2}\mathbf{k}_{2}}{(2\pi )^{2}}\left\vert 
{\rho_{-\mathbf{k}_{2}}\left( t\right) }\right\vert
^{2}\left\vert E_{\mathbf{k}_{2}}(t)\right\vert ^{2}\sin ^{2}\theta _{2},
\end{equation}
where $\mathcal{J}=\int_{0}^{2\pi}{d\theta}/({1+{\cos^2{\theta}}})\simeq 4.4.$ The radiation rates scale as $\dot{\eta}^{\pm }_{2D}\propto \left( v_{T}/c\right)
^{2}=c_{L}^{-2}$ and $\dot{\eta}^{\pm }_{2D}\propto \Delta N,$ in agreement with our simulation results. 

\section{Conclusion}

Whereas several electromagnetic radiation mechanisms at the plasma frequency have been proposed during last decades, the linear mode conversion process at constant frequency (LMC) has been shown to be dominant in plasmas with random density fluctuations as the solar wind. This work presents a new theoretical and numerical model which describes in two-dimensional geometry all possible linear interactions between upper-hybrid wave turbulence and random density fluctuations in a weakly magnetized  and inhomogeneous plasma; not only linear processes as wave reflection, refraction, scattering, tunneling, trapping, or mode conversion are taken into account, but also linear wave coupling, interferences between scattered waves, etc. The model describes interactions between wave and density turbulence as close as possible to reality. 

The model considers a radio source as a weakly magnetized plasma where random density fluctuations and  upper-hybrid wavepackets evolve from initial spectra according to modified Zakharov equations including weak magnetic effects. The current generated by the interactions of turbulent upper-hybrid wavepackets with density fluctuations  radiates electromagnetic waves by linear mode conversion at constant frequency; those are leaving the randomly inhomogeneous source and propagate freely in an external homogeneous plasma. Such process is possible due to upper-hybrid waves' trapping in plasma density depletions.

Compact equations describing the time evolution of electric and magnetic fields radiated in the $\mathcal{O}$, $\mathcal{X}$ and $\mathcal{Z}$ modes by the current, as well as the dispersion and polarization properties of modes, are obtained analytically and solved numerically, providing the time variations of electromagnetic energies and corresponding radiation rates. Jointly, on the basis of the numerical results that validate theoretical hypotheses, analytical calculations are conducted in 3D geometry in the framework of weak turbulence theory extended to randomly inhomogeneous plasmas, that recover the main physical conclusions stated using the new model. 

In a first step, electromagnetic radiation is studied in unmagnetized plasmas, where only the ordinary mode $\mathcal{O}$ exists. Then, the work is extended to weakly magnetized plasmas and to the determination of electromagnetic radiation rates and energies of $\mathcal{O}$, $\mathcal{X}$ and $\mathcal{Z}$ modes. Their dependencies with plasma parameters as the magnetization ratio $\omega_c/\omega_p$, the  electron thermal velocity ratio $v_T/c$ and the average level of random density fluctuations $\Delta N$ is determined in the form of scaling laws. In particular, this study is conducted for various initial upper-hybrid wave and density spectra and completed by the analytic determination of radiation rates for any mode and for any given wave and density spectra.

This work opens a new way to analyze the efficiency of electromagnetic emissions at plasma frequency by realistic wave and density turbulence spectra interacting in weakly magnetized solar wind plasmas.

\section{acknowledgements} 	
		This work was granted access to the HPC computing and storage resources under the allocation 2023-A0130510106 and 2024-A017051010 made by GENCI. This research was also financed in part by the French National Research Agency (ANR) under the project ANR-23-CE30-0049-01.  C.K. thanks the International Space Science Institute (ISSI) in Bern through ISSI International Team project No. 557, Beam-Plasma Interaction in the Solar Wind and the Generation of Type III Radio Bursts. C. K. thanks the Institut Universitaire de France (IUF).
        
        For open access purposes, a CC-BY license has been applied by the authors to this document and will be applied to any subsequent version up to the author's manuscript accepted for publication resulting from this submission.

\begin{appendix}
\section{Non-potential upper-hybrid waves}\label{sec:app}        
Let us determine the dispersion relation of non-potential upper-hybrid waves. Note that these waves are also named in other works as Langmuir/$\mathcal{Z}$-mode waves (e.g. \cite{Bale1996}, \cite{GrahamCairns2013}, 
\cite{Kellogg2013}) or $\mathcal{LZ}$  waves (\cite{Polanco2025a}, \cite{Krafft2025}). 
In a homogeneous magnetized plasma, the current $\delta 
\mathbf{J=-}en_{0}\mathbf{v}_{e}$ due to the  motion of electrons of density $n_0$, velocity  $\mathbf{v}_e$ and charge $-e<0$ contains the non-potential part 
\begin{equation}
\delta \mathbf{J}_{np}=\frac{1}{4\pi}\frac{\omega _{p}^{2}\omega _{c}}{\omega
^{2}-\omega _{c}^{2}}\mathbf{h}\times \mathbf{\nabla }_{\perp }\varphi , 
  \label{A1}
\end{equation}
which satisfies $\mathbf{\nabla }\cdot \delta \mathbf{J}_{np}=0;$ $\omega$, $\omega_c$, and $\omega_p$ are the wave, cyclotron and plasma frequencies; $\mathbf{h}=\mathbf{B}_0/B_0; $ $\varphi $ is the wave potential. According
to Maxwell equations, a non-potential electric field $\delta \mathbf{E}$ ($%
\nabla \cdot \delta \mathbf{E}=0$) appears as
\begin{equation}
\nabla \times \delta \mathbf{B}\simeq \frac{c\nabla \times \left( \nabla
\times \delta \mathbf{E}\right) }{i\omega }\simeq -\frac{c\nabla ^{2}\delta 
\mathbf{E}}{i\omega }+\frac{c\nabla \cdot \delta \mathbf{E}}{i\omega } \\ 
\simeq-\frac{c\nabla ^{2}\delta \mathbf{E}}{i\omega }\simeq \frac{4\pi }{c}\delta 
\mathbf{J}_{np},\label{A2}
\end{equation}
where the displacement current is neglected. Using the Coulomb gauge, we can
write that $\nabla \times \delta \mathbf{B}=\nabla \times \left( \nabla \times \delta 
\mathbf{A}\right) =-\nabla ^{2}\delta \mathbf{A}={4\pi}/{c\delta 
\mathbf{J}_{np}},$ so that the non-potential part of the electric field is 
\begin{equation}
\delta \mathbf{E}\simeq \frac{i\omega }{c}\delta \mathbf{A}\simeq -\frac{%
i\omega \omega _{p}^{2}\omega _{c}}{c^{2}\left( \omega ^{2}-\omega
_{c}^{2}\right) }\nabla ^{-2}\left( \mathbf{h}\times \mathbf{\nabla }_{\perp
}\varphi \right) .\label{A4}
\end{equation}
Then we can estimate the ratio (see also \ref{l1}) 
\begin{equation}
\frac{\left\vert \delta \mathbf{E}\right\vert }{\left\vert \mathbf{\nabla }%
\varphi \right\vert }\sim \frac{\omega _{p}^{2}}{c^{2}k^{2}}\frac{\omega
\omega _{c}}{\left( \omega ^{2}-\omega _{c}^{2}\right) }.  \label{A5}
\end{equation}
Note that if $k_{\parallel }\ll k,$ one can derive from (\ref{A2}) that $\delta \mathbf{B\simeq }-{\omega _{p}^{2}\omega _{c}}\mathbf{h}\varphi /{c( \omega
^{2}-\omega _{c}^{2})}.$ Then, adding the non-potential electric field as a correction into the electron velocity (\ref{ve1}), we get in a weakly magnetized plasma that 

\begin{equation}
\mathbf{v}_{e}\simeq -\frac{i}{\omega }\frac{e}{m_{e}}\mathbf{E}+\frac{e}{%
m_{e}}\frac{\omega _{c}}{\omega ^{2}-\omega _{c}^{2}}\left( \mathbf{h}\times 
\mathbf{E}_{\perp }\right) 
-\frac{e}{m_{e}c}\nabla ^{-2}\frac{\omega
_{p}^{2}\omega _{c}}{c\left( \omega ^{2}-\omega _{c}^{2}\right) }\mathbf{h}%
\times \mathbf{\nabla }_{\perp }\varphi ,\label{A7}
\end{equation}

where $m_e$ is the electron mass. The density corresponding to the non-potential part of the velocity is 
\begin{equation}
4\pi e\delta n_{e}=\frac{\omega _{p}^{2}}{i\omega }\frac{\omega _{c}}{\omega
^{2}-\omega _{c}^{2}}\nabla \cdot \left( \mathbf{h}\times \delta \mathbf{E}%
\right) \simeq \frac{\omega _{p}^{4}\omega _{c}^{2}}{c^{2}\left( \omega
^{2}-\omega _{c}^{2}\right) ^{2}}\varphi. \label{A8}
\end{equation}
Using the Poisson equation $\nabla \cdot \hat{\varepsilon}\ \mathbf{E}=-4\pi e\delta n_{e}$,
where $\hat{\varepsilon}$ is the dielectric constant of upper-hybrid
waves (neglecting the ions' contribution) 
\begin{equation}
\hat{\varepsilon}\left( \omega ,k\right) =1-\frac{\omega _{p}^{2}}{\omega
^{2}-\omega _{c}^{2}}\frac{k_{\bot }^{2}}{k^{2}}-\frac{\omega _{p}^{2}}{%
\omega ^{2}}\frac{k_{\parallel }^{2}}{k^{2}},\label{A9}
\end{equation}
we get 
\begin{equation}
\nabla ^{2}\left( \nabla \cdot \ \hat{\varepsilon}\ \nabla \varphi \right)
\simeq \left( \frac{\omega _{p}^{2}\omega _{c}}{c\left( \omega ^{2}-\omega
_{c}^{2}\right) }\right) ^{2}\nabla _{\perp }^{2}\varphi  \label{A10}
\end{equation}
and, in the Fourier space
\begin{equation}
\left( \hat{\varepsilon}_{k}+\frac{\omega _{p}^{2}}{c^{2}k^{2}}\left( \frac{%
\omega _{p}\omega _{c}}{\omega ^{2}-\omega _{c}^{2}}\right) ^{2}\frac{%
k_{\perp }^{2}}{k^{2}}\right) \varphi _{k}=\hat{\varepsilon}_{uh}\varphi
_{k}=0,\label{A11}
\end{equation}
where $\hat{\varepsilon}_{uh}$ is the effective dielectric constant of
weakly non-potential upper-hybrid waves in a weakly magnetized plasma,
corresponding to the dispersion (with added thermal effects and $\omega_c/\omega_p\ll1$) 
\begin{equation}
\omega \simeq \omega _{p}+\frac{3}{2}\omega _{p}\left( k\lambda _{D}\right)
^{2}+\frac{\omega _{c}^{2}}{2\omega _{p}}\sin^2 \theta\left(
1-\frac{\omega _{p}^{2}}{c^{2}k^{2}}\right) ,  \label{uh-dispes2}
\end{equation}
where $\sin^2 \theta=k_{\perp}^2/k^2$; $\omega_{p}^{2}/{c^{2}k^{2}}\ll1$ can be neglected in the potential (electrostatic) limit.

\section{Dispersion of electromagnetic waves in the vicinity of
cutoff frequencies in a weakly magnetized homogeneous plasma}\label{sec:app}
This appendix is devoted to derive, in a cold and homogeneous plasma, the dispersion relations of $\mathcal{O}$, $\mathcal{X}$ and $\mathcal{Z}$-mode waves near their cutoff frequencies. Those will be used to determine the lhs terms of equations as (\ref{dbkdt}) and (\ref{xznorm}), where external density fluctuations only appear in the rhs terms. 

In the theoretical model presented in the main text, the
ambient magnetic field $\mathbf{B}_{0}$ is directed along the $x$-axis,
which is chosen as the parallel direction; the perpendicular plane is
defined by $(y,z)$. In the calculations shown below, we use the most common
frame for readers, where $\mathbf{B}_{0}$ is directed along the $z$-axis;
the variables with the subscript $"\parallel
" $ indicate parallel propagation. The axes $x$ and $y$ of the
perpendicular plane are indicated by the subscripts $\perp $ and $\perp
^{\prime }$, respectively. As usually done, we choose below a reference frame where $k_{\perp ^{\prime }}=0.$

In a cold magnetized plasma, the Maxwell equations provide the following
relations 
\begin{equation}
\left( \varepsilon _{\perp }-\frac{c^{2}k_{\parallel }^{2}}{\omega ^{2}}%
\right) E_{\perp }+iqE_{\perp ^{\prime }}=-\frac{c^{2}k_{\perp }k_{\parallel
}}{\omega ^{2}}E_{\parallel },  \label{p1}
\end{equation}
\begin{equation}
-iqE_{\perp }+\left( \varepsilon _{\perp }-\frac{c^{2}k^{2}}{\omega ^{2}}%
\right) E_{\perp ^{\prime }}=0,  \label{p2}
\end{equation}
\begin{equation}
\left( \varepsilon _{\Vert }-\frac{c^{2}k_{\perp }^{2}}{\omega ^{2}}\right)
E_{\parallel }=-\frac{c^{2}k_{\perp }k_{\parallel }}{\omega ^{2}}E_{\perp },
\label{p3}
\end{equation}
where  $\varepsilon _{\perp }=\varepsilon _{\perp ^{\prime }}=1-{\omega _{p}^{2}%
}/({\omega ^{2}-\omega _{c}^{2}})$,  $\varepsilon _{\Vert }=1-{\omega _{p}^{2}}/{\omega ^{2}}$ and $q=-{\omega _{c}\omega_{p}^{2}}/{\omega \left( \omega ^{2}-\omega _{c}^{2}\right) }$
are the matrix elements of the dielectric tensor 
\begin{equation}
\hat{\varepsilon}\left( \omega ,k\right) =\left[ 
\begin{array}{ccc}
\varepsilon _{\perp } & iq & 0 \\ 
-iq & \varepsilon _{\perp } & 0 \\ 
0 & 0 & \varepsilon _{\Vert }
\end{array}%
\right] . \label{B4}
\end{equation}
From equations (\ref{p1})-(\ref{p3}) we obtain that
\begin{equation}
\left[ \left( \frac{k_{\parallel }^{2}c^{2}}{\omega ^{2}}-\varepsilon
_{\perp }\right) \left( \frac{k^{2}c^{2}}{\omega ^{2}}-\varepsilon _{\perp
}\right) -q^{2}\right] \left( \frac{k_{\perp }^{2}c^{2}}{\omega ^{2}}%
-\varepsilon _{\Vert }\right) -\frac{k_{\parallel }^{2}k_{\perp
}^{2}c^{4}}{\omega ^{4}}\left( \frac{k^{2}c^{2}}{\omega ^{2}}%
-\varepsilon _{\perp }\right) =0,  \label{Lin2}
\end{equation}
which determines the linear wave dispersion $\omega =\omega \left( 
\mathbf{k}\right) $ in a cold plasma. Note that terms proportional to $k^{6}$ cancel. Introducing the squared refractive
indices $N^{2}=c^{2}k^{2}/\omega ^{2}$ and $N_{\parallel ,\perp
}^{2}=c^{2}k_{\parallel ,\perp }^{2}/\omega ^{2}$, as well as the propagation
angle $\theta $ with respect to $\mathbf{B}_{0}$ ($\mathbf{k}\cdot\mathbf{B}_0=kB_0\cos\theta$), we get the
biquadratic equation providing the dependence $k^{2}( \omega ,\theta
) $ (see also \cite{Shafranov1967})
\begin{equation}
N^{4}\left( \varepsilon _{\perp }\sin ^{2}\theta +\varepsilon _{\parallel
}\cos ^{2}\theta \right) \\
+N^{2}\left( q^{2}\sin ^{2}\theta -\varepsilon
_{\perp }\left( \varepsilon _{\perp }\sin ^{2}\theta +\varepsilon
_{\parallel }\cos ^{2}\theta \right) -\varepsilon _{\perp }\varepsilon
_{\parallel }\right) 
-q^{2}\varepsilon _{\parallel }+\varepsilon _{\perp
}^{2}\varepsilon _{\parallel }=0.  \label{Shafranov}
\end{equation}

\subsection{Dispersion near the $\mathcal{O}$-mode cutoff frequency}
The ordinary mode $\mathcal{O}$ cannot propagate strictly parallel to the ambient
magnetic field but oscillates near $\omega _{p}$. However, for perpendicular
propagation ($k_{\parallel }=0$), equation (\ref{p3}) can be separated from
 (\ref{p1})-(\ref{p2}) and  $\mathcal{O}$-mode waves follow the
dispersion relation $\omega ^{2}=\omega _{p}^{2}+c^{2}k_{\perp }^{2}$, i.e.  $\omega\simeq\omega _{p}+c^{2}k^{2}\sin^{2}{\theta}/2\omega_p$.
In this case, non vanishing wave electric and magnetic field components are
$E_{\parallel }$ and $B_{\perp ^{\prime }}=-\left( ck_{\perp }/\omega \right)
E_{\parallel }$.
In the case of oblique propagation, the dispersion of $\mathcal{O}$-mode
waves can be separated from those of $\mathcal{X}$ and $\mathcal{Z}$-mode waves when $E_{\parallel }$\
is the dominant electric field component and $B_{\parallel }$ is small or vanishing. However, even in a weakly magnetized plasma with $\omega _{c}\ll
\omega _{p},$ it is not possible to neglect the other field components $%
B_{\perp }=-\left( ck_{\parallel }/\omega \right) E_{\perp ^{\prime }}$, $%
B_{\perp ^{\prime }}=\left( ck_{\parallel }/\omega \right) E_{\perp }-\left(
ck_{\perp }/\omega \right) E_{\parallel },$ and $B_{\parallel }=\left(
ck_{\perp }/\omega \right) E_{\perp ^{\prime }},$ with $\nabla \cdot \mathbf{%
B}=k_{\perp }B_{\perp }+k_{\parallel }B_{\parallel }=0$ $(k_{\perp ^{\prime
}}=0)$. 

\begin{figure}[H]
    \centering
    \includegraphics[width=0.5\textwidth]{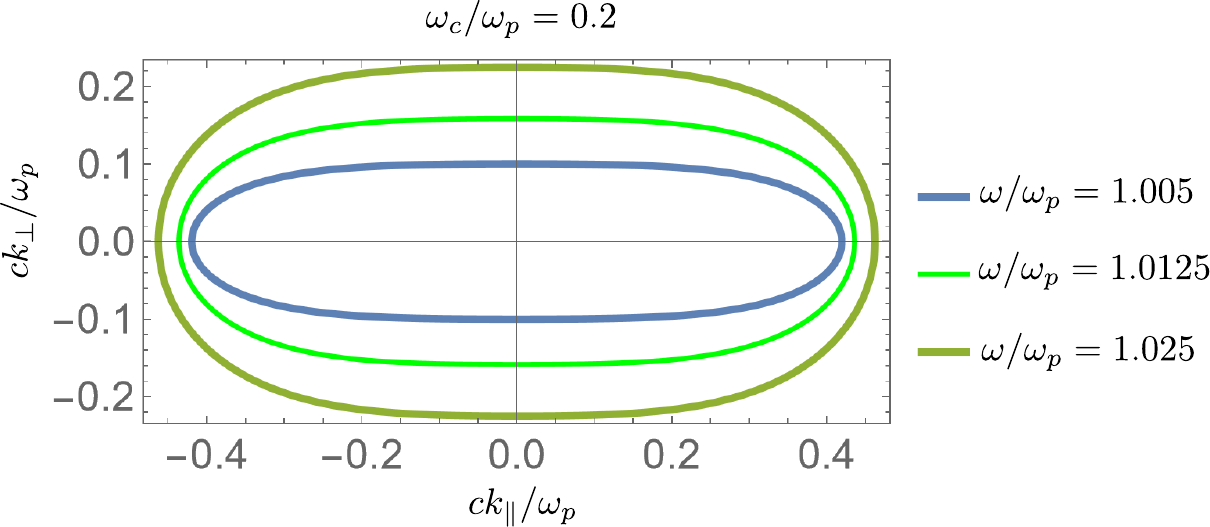}
    \caption{Electromagnetic $\mathcal{O}$-mode  waves : dispersion curves in the map $( ck_{\parallel }/\omega _{p},  ck_{\perp }/\omega_{p}) $, for $
\omega _{c}/\omega _{p}=0.2$  and the frequencies $\omega /\omega _{p}=1.005,$ $1.0125,$ $1.025$ (see legend).}
    \label{fig15}
\end{figure}

Fig. \ref{fig15} shows the exact solutions of equation (\ref{Shafranov}) in the form
of isocontours $\omega \left( k_{\parallel },k_{\perp }\right)=cst$ in the map $(k_\parallel, k_\perp)$ ,  for $
\omega _{c}/\omega _{p}=0.2.$ One can see that the magnetic field is responsible for the elongation (anisotropy) along the parallel direction of the ordinary wave dispersion. Note that for $\omega_c=0$, the  isocontours  $\omega \left( k_{\parallel },k_{\perp }\right)=cst$  are circular. For a weaker magnetic field with $\omega _{p}^{2}\gg c^{2}k_{\perp
}^{2}$ and $c^{2}k_{\parallel }^{2}\geq \omega _{c}\omega _{p},$ it is problematic
to separate the $\mathcal{O}$-mode from the $\mathcal{X}$- and $\mathcal{Z}$-modes; however, in this case, it is possible to consider an unmagnetized plasma.

For analytic calculation purposes, the dispersion relation of $\mathcal{O}$-mode waves in a weakly magnetized plasma can be approximated near their cutoff frequency by  
\begin{equation}
\omega_k \simeq \omega _{p}+\frac{k^{2}c^{2}}{2\omega _{p}}\sin ^{2}\theta \label{disp1}
\end{equation}
for $k^{2}c^{2}\sin ^{2}\theta \leq \omega _{p}\omega _{c}$ ($\sin^{2}\theta \neq 0$), and by
\begin{equation}
\omega_k \simeq \omega _{p}+\frac{\omega _{c}^{2}-\omega _{p}\omega _{c}\cos
^{2}\theta }{2\omega _{p}}+\frac{k^{2}c^{2}}{2\omega _{p}}   \label{disp2}
\end{equation}
for larger wavenumbers and wave frequencies. The  formula (\ref{disp2}) at $k=0$ corresponds to the exact dispersion of $\mathcal{O}$-mode waves at their cutoff frequency;  the corrective term $c^2k^2/2\omega_p$ is added  to describe ordinary mode wave propagation at very small $k$. These approximations result in relative errors ranging from 1 to $10\%$, depending on $k$, $\theta$ and $\omega_c$, as revealed by numerical studies (not shown here). They  are only used to perform analytic calculations and avoid the singularity that exists in the region where  both $\mathcal{O}$ and $\mathcal{Z}$ modes' dispersion curves meet when the angle of propagation tends to zero.

\begin{figure}[H]
    \centering
    \includegraphics[width=0.4\textwidth]{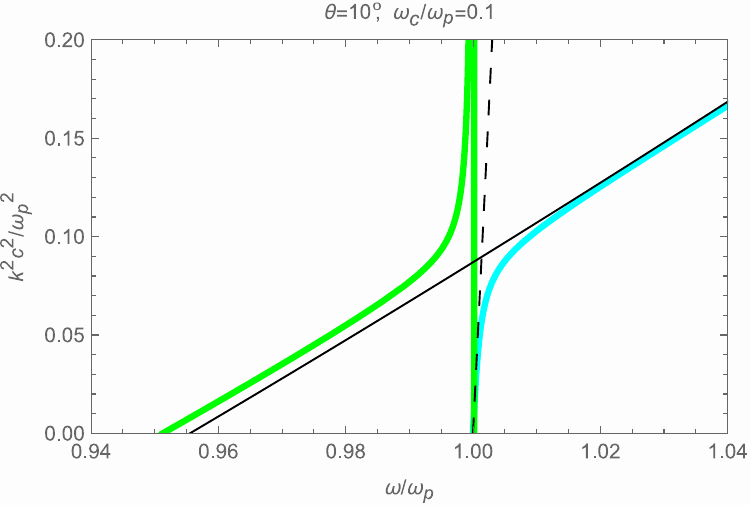}
    \caption{Dispersion of electromagnetic modes near the frequency $\omega_p$, for $\omega _{c}/\omega _{p}=0.1$  and $\theta=10^{\circ}$ : variation of $c^2k^2/\omega^2_p$ as a function of $\omega /\omega _{p}$. Green (blue) lines represent the $\mathcal{Z}$-mode waves and the curve $\omega=\omega_p$ (the
    $\mathcal{O}$-mode waves). Black dashed and solid lines represent the dispersion curves (\ref{disp1}) and (\ref{disp2}) near $\omega\simeq\omega_p$, respectively. }
    \label{fig16}
\end{figure}

Fig. \ref{fig16} shows the exact solution of equation (\ref{Shafranov}) in
the map  $(\omega/\omega _{p},  c^2k^2/{\omega^2_{p}})$,  for $\omega _{c}/\omega _{p}=0.1$  and $\theta=10^{\circ}$. Green and blue lines represent  the $\mathcal{Z}$-mode waves and the curve $\omega=\omega_p$, as well as the $\mathcal{O}$-mode waves, respectively. The solid and dashed black lines, which fit with good accuracy the $\mathcal{O}$-mode wave dispersion from $\omega /\omega _{p}\simeq1.015$ to  $\omega /\omega _{p}\simeq1.04$ and near $\omega /\omega _{p}\simeq1$, respectively, represent the dispersion curves (\ref{disp2}) and (\ref{disp1}). 

\subsection{Dispersion near the $\mathcal{X}$- and $\mathcal{Z}$-modes'
cutoff frequencies}

\subsubsection{Parallel propagation}
In the parallel propagation case ($k_{\perp}=0$), the fields of  $\mathcal{X}$- and $\mathcal{Z}$-mode waves
satisfy $B_{\perp }=-\left( ck_{\parallel }/\omega \right) E_{\perp ^{\prime
}},\ B_{\perp ^{\prime }}=\left( ck_{\parallel }/\omega \right) E_{\perp },\ 
$and$\ \ B_{\parallel }=0,$ with $k_{\perp }B_{\perp }+k_{\parallel
}B_{\parallel }=0$ ($\nabla \cdot \mathbf{B}=0,$ $k_{\perp ^{\prime }}=0$). Introducing the perpendicular electric field in the form $E_{\pm }=E_{\perp}\pm iE_{\perp^{\prime}}$, 
we get from (\ref{p1})-(\ref{p3}) that $( \varepsilon _{\pm
}-c^{2}k_{\parallel }^{2}/\omega ^{2}) E_{\pm }=0$, with the
dispersion relation
\begin{equation}
\varepsilon _{\pm }=\varepsilon _{\perp }\pm q=1-\frac{\omega _{p}^{2}}{%
\omega \left( \omega \mp \omega _{c}\right) }=\frac{c^{2}k_{\parallel }^{2}}{%
\omega ^{2}}.  \label{parr}
\end{equation}
Moreover, the relation $\varepsilon _{\Vert }E_{\parallel }=0$ (\ref{p3}),
together with $\varepsilon _{\Vert }\neq 0,$ leads to $E_{\parallel }=0$.
Note that the field $E_{+}$ ($E_{-}$) corresponds to the $\mathcal{X}$-mode ($\mathcal{Z}$-mode), with $E_{-}=0$ ($E_{+}=0$) and the cutoff frequencies $\omega _{\pm }=({\omega _{p}^{2}+{\omega _{c}^2}/4})^{1/2}\pm \omega _{c}/2$  provided by $\varepsilon _{\pm}=0$. 
Then, for parallel propagation ($k_{\perp }=0$), the dispersion relation
near the cutoff $k_{\parallel} \simeq 0 $ can be calculated for both
modes as 
\begin{equation}
2\omega _{\pm }\delta \omega =c^{2}k_{\parallel}^{2}\frac{\left( \omega _{\pm }\mp
\omega _{c}\right) }{\omega _{\pm }},\omega _{k}^{\pm
}=\omega _{\pm }+\delta \omega \simeq \omega _{\pm }+\frac{c^{2}k_{\parallel
}^{2}}{2\omega _{\pm }}\left( 1\mp \frac{\omega _{c}}{\omega _{\pm }}\right)
  \label{parXZ}
\end{equation}
where we neglected very small terms as $( \delta \omega)^{2},$ $\omega _{c}\delta \omega ,$  and $c^{2}k_{\parallel}^{2}\delta \omega .$

\subsubsection{Perpendicular propagation}
In the perpendicular propagation case ($k_{\parallel }=0$), equations (%
\ref{p1})-(\ref{p3}) lead to $\left( \varepsilon _{\perp }-q\right) \left(
\varepsilon _{\perp }+q\right) =N_{\perp }^{2}\varepsilon _{\perp }.$
Supposing that $N_{\perp }^{2}\ $is very small, we get for the $\mathcal{Z}$%
-mode that $\varepsilon _{\perp }-q\simeq 0$ and 
\begin{equation}
\varepsilon _{-}=\varepsilon _{\perp }-q=N_{\perp }^{2}\frac{\varepsilon
_{\perp }}{\varepsilon _{\perp }+q}\simeq \frac{1}{2}N_{\perp }^{2}=\frac{%
c^{2}k_{\perp }^{2}}{2\omega ^{2}},\label{B6}
\end{equation}
providing the dispersion relation%
\begin{equation}
\omega \left( \omega +\omega _{c}\right) -\omega _{p}^{2}\simeq \frac{%
c^{2}k_{\perp }^{2}}{2\omega }\left( \omega +\omega _{c}\right) . \label{B7}
\end{equation}
Similarly, we get for the $\mathcal{X}$-mode that%
\begin{equation}
\omega \left( \omega -\omega _{c}\right) -\omega _{p}^{2}=N_{\perp }^{2}%
\frac{\varepsilon _{\perp }}{\varepsilon _{\perp }-q}\simeq \frac{%
c^{2}k_{\perp }^{2}}{2\omega }\left( \omega -\omega _{c}\right) .\label{B8}
\end{equation}
Introducing the cutoff frequencies $\omega _{\pm },$ we obtain finally
\begin{equation}
\omega _{k}^{\pm }\simeq \omega _{\pm }+\frac{c^{2}k_{\perp }^{2}}{4\omega
_{\pm }}\left( 1\mp \frac{\omega _{c}}{\omega _{\pm }}\right) . 
\label{perpXZ}
\end{equation}

\subsubsection{Oblique propagation with $E_{\parallel }=0$}
In the oblique propagation case, $E_{\parallel }\neq 0$. However, let us
first study the case when $E_{\parallel }$ can be neglected. Using that $E_{\pm }=E_{\perp}\pm iE_{\perp^{\prime}}$, i.e. $2E_{\perp }=E_{+}+E_{-}$ and $2E_{\perp ^{\prime
}}=-i\left( E_{+}-E_{-}\right) ,$ equations (\ref{p1})-(\ref{p3}) lead to%
\begin{eqnarray}
2\left( \varepsilon _{\perp }+q\right) E_{+} &=&\left( \frac{%
c^{2}k_{\parallel }^{2}}{\omega ^{2}}+\frac{c^{2}k^{2}}{\omega ^{2}}\right)
E_{+}-\frac{c^{2}k_{\perp }^{2}}{\omega ^{2}}E_{-}, \\
2\left( \varepsilon _{\perp }-q\right) E_{-} &=&-\frac{c^{2}k_{\perp }^{2}}{%
\omega ^{2}}E_{+}+\left( \frac{c^{2}k^{2}}{\omega ^{2}}+\frac{%
c^{2}k_{\parallel }^{2}}{\omega ^{2}}\right) E_{-}.\label{B10}
\end{eqnarray}
Then, assuming that $E_{-}\rightarrow 0$ ($E_{+}\rightarrow 0$) for the $\mathcal{X}$-mode ($\mathcal{Z}$-mode), we obtain that 
\begin{equation}
\varepsilon _{+}E_{+}=\frac{c^{2}k_{\parallel }^{2}+c^{2}k^{2}}{2\omega ^{2}}%
E_{+},\text{ \ \ \ \ \ \ \ }\varepsilon _{-}E_{-}=\frac{c^{2}k_{\parallel
}^{2}+c^{2}k^{2}}{2\omega ^{2}}E_{-},\label{B11}
\end{equation}
where $\varepsilon _{\pm }=\varepsilon _{\perp }\pm q.$ When $\omega $ is
close to the cutoff frequencies $\omega _{\pm }$, we can write that 
\begin{equation}
\varepsilon _{\pm }=\frac{\omega \left( \omega \mp \omega _{c}\right)
-\omega _{p}^{2}}{\omega \left( \omega \mp \omega _{c}\right) }\simeq \frac{%
2\omega _{\pm }\left( \omega -\omega _{\pm }\right) }{\omega _{\pm }\left(
\omega _{\pm }\mp \omega _{c}\right) },\label{B12}
\end{equation}
so that, for $\mathcal{X}$- and $\mathcal{Z}$-modes, we get the dispersion
laws for oblique propagation near the cutoff frequencies $\omega _{\pm }$ as
\begin{equation}
\omega _{k}^{\pm }\simeq \omega _{\pm }+\frac{c^{2}k_{\parallel
}^{2}+c^{2}k^{2}}{4\omega _{\pm }}\left( 1\mp \frac{\omega _{c}}{\omega
_{\pm }}\right) .  \label{oblique1}
\end{equation}
As expected, we recover  (\ref{parXZ}) and (\ref{perpXZ}) when $k_{\perp
}=0$ and $k_{\parallel }=0$, respectively.

\subsubsection{Oblique propagation with $E_{\parallel }\neq 0$}

Let us now assume that $E_{\parallel }\ $is non vanishing but small. In
this case, equations (\ref{p1})-(\ref{p3}) lead to%
\begin{equation}
\left( \varepsilon _{\perp }+q-\frac{c^{2}k_{\parallel }^{2}}{\omega ^{2}}%
\right) E_{+}+\left( \varepsilon _{\perp }-q-\frac{c^{2}k_{\parallel }^{2}}{%
\omega ^{2}}\right) E_{-}  = 
\frac{c^{2}k_{\parallel }^{2}}{\omega ^{2}}\frac{%
c^{2}k_{\perp }^{2}}{\omega ^{2}-\omega _{p}^{2}-c^{2}k_{\perp }^{2}}\frac{\left(
E_{+}+E_{-}\right)}{2} ,\label{B14}
\end{equation}%
\begin{equation}
\left( \varepsilon _{\perp }+q\right) E_{+}+\left( \varepsilon _{\perp
}-q\right) E_{-}=\frac{c^{2}k_{\parallel }^{2}}{\omega ^{2}}\left( 1+\frac{%
c^{2}k_{\perp }^{2}}{\omega ^{2}-\omega _{p}^{2}-c^{2}k_{\perp }^{2}}\right)
\left( E_{+}+E_{-}\right) ,  \label{epsplus}
\end{equation}%
\begin{equation}
\left( \varepsilon _{\perp }+q\right) E_{+}-\left( \varepsilon _{\perp
}-q\right) E_{-}=\frac{c^{2}k^{2}}{\omega ^{2}}\left( E_{+}-E_{-}\right) .
\label{XZdispersion}
\end{equation}%
One observes that terms of spatial dispersion couple together the $\mathcal{X}$-
and $\mathcal{Z}$-modes. If this coupling is weak, as can be supposed near
the cutoff $k\simeq 0$, $\mathcal{X}$- and $\mathcal{Z}$-modes can be separated by
neglecting $E_{-}$ and $E_{+}$ in equations (\ref{epsplus}) and (\ref%
{XZdispersion}), respectively; then we get 
\begin{equation}
2\left( \varepsilon _{\perp }+q\right) E_{+}\simeq \left[ \frac{c^{2}k^{2}}{%
\omega ^{2}}+\frac{c^{2}k_{\parallel }^{2}}{\omega ^{2}}\left( 1+\frac{%
c^{2}k_{\perp }^{2}}{\omega ^{2}-\omega _{p}^{2}-c^{2}k_{\perp }^{2}}\right) %
\right] \left( E_{+}+E_{-}\right),  \label{xmod}
\end{equation}%
\begin{equation}
2\left( \varepsilon _{\perp }-q\right) E_{-}\simeq \left[ \frac{%
c^{2}k_{\parallel }^{2}}{\omega ^{2}}\left( 1+\frac{c^{2}k_{\perp }^{2}}{%
\omega ^{2}-\omega _{p}^{2}-c^{2}k_{\perp }^{2}}\right) -\frac{c^{2}k^{2}}{%
\omega ^{2}}\right] \left( E_{+}+E_{-}\right).  \label{zmod}
\end{equation}%
Thus, the correction provided to dispersion by the non vanishing field $%
E_{\parallel }\neq 0$ amounts to multiplying the term ${c^{2}k_{\parallel
}^{2}}/{\omega _{p}^{2}}$ in equation (\ref{oblique1}) by 
$(1+{c^{2}k_{\perp}^{2}}/{(\mathrm{\pm }\omega _{c}\omega _{p}-c^{2}k_{\perp }^{2}})) $.
Note that if $c^{2}k_{\perp }^{2}\leq $ $\omega _{c}\omega _{p}$, this
correction is not essential.
      
	\end{appendix}

\bibliography{./articleBiblio.bib}

\begin{thebibliography}{}
\expandafter\ifx\csname natexlab\endcsname\relax\def\natexlab#1{#1}\fi
\providecommand{\url}[1]{\href{#1}{#1}}
\providecommand{\dodoi}[1]{doi:~\href{http://doi.org/#1}{\nolinkurl{#1}}}
\providecommand{\doeprint}[1]{\href{http://ascl.net/#1}{\nolinkurl{http://ascl.net/#1}}}
\providecommand{\doarXiv}[1]{\href{https://arxiv.org/abs/#1}{\nolinkurl{https://arxiv.org/abs/#1}}}

\bibitem[{{Akimoto}(1989)}]{Akimoto1989}
{Akimoto}, K. 1989, Physics of Fluids B, 1, 1998, \dodoi{10.1063/1.859063}

\bibitem[{{Badman} {et~al.}(2022){Badman}, {Carley}, {Ca{\~n}izares}, {Dresing}, {Jian}, {Lario}, {Gallagher}, {Mart{\'\i}nez Oliveros}, {Pulupa}, \& {Bale}}]{Badman2022}
{Badman}, S.~T., {Carley}, E., {Ca{\~n}izares}, L.~A., {et~al.} 2022, \apj, 938, 95, \dodoi{10.3847/1538-4357/ac90c2}

\bibitem[{{Bale} {et~al.}(1996){Bale}, {Burgess}, {Kellogg}, {Goetz}, {Howard}, \& {Monson}}]{Bale1996}
{Bale}, S.~D., {Burgess}, D., {Kellogg}, P.~J., {et~al.} 1996, \grl, 23, 109, \dodoi{10.1029/95GL03595}

\bibitem[{Cairns \& Layden(2018)}]{CairnsLayden2018}
Cairns, I.~H., \& Layden, A. 2018, Physics of Plasmas, 25, 082309, \dodoi{10.1063/1.5037300}

\bibitem[{{Cairns} \& {Willes}(2005)}]{CairnsWilles2005}
{Cairns}, I.~H., \& {Willes}, A.~J. 2005, Physics of Plasmas, 12, 052315, \dodoi{10.1063/1.1889123}

\bibitem[{{Celnikier} {et~al.}(1983){Celnikier}, {Harvey}, {Jegou}, {Moricet}, \& {Kemp}}]{Celnikier1983}
{Celnikier}, L.~M., {Harvey}, C.~C., {Jegou}, R., {Moricet}, P., \& {Kemp}, M. 1983, \aap, 126, 293

\bibitem[{Chen {et~al.}(2021)Chen, Ma, Wu, Zhao, Tang, \& Bale}]{Chen2021}
Chen, L., Ma, B., Wu, D., {et~al.} 2021, \apjl, 915, L22, \dodoi{10.3847/2041-8213/ac0b43}

\bibitem[{{Dulk}(1985)}]{Dulk1985}
{Dulk}, G.~A. 1985, \araa, 23, 169, \dodoi{10.1146/annurev.aa.23.090185.001125}

\bibitem[{{Dum} \& {Nishikawa}(1994)}]{DumNishikawa1994}
{Dum}, C.~T., \& {Nishikawa}, K.~I. 1994, Physics of Plasmas, 1, 1821, \dodoi{10.1063/1.870636}

\bibitem[{Edney \& Robinson(1999)}]{EdneyRobinson1999}
Edney, S.~D., \& Robinson, P.~A. 1999, Phys. Plasmas, 6, 3799, \dodoi{10.1063/1.873644}

\bibitem[{{Fox} {et~al.}(2016){Fox}, {Velli}, {Bale}, {Decker}, {Driesman}, {Howard}, {Kasper}, {Kinnison}, {Kusterer}, {Lario}, {Lockwood}, {McComas}, {Raouafi}, \& {Szabo}}]{Fox2016}
{Fox}, N.~J., {Velli}, M.~C., {Bale}, S.~D., {et~al.} 2016, \ssr, 204, 7, \dodoi{10.1007/s11214-015-0211-6}

\bibitem[{{Ginzburg} \& {Zhelezniakov}(1958)}]{GinzburgZheleznyakov1958}
{Ginzburg}, V.~L., \& {Zhelezniakov}, V.~V. 1958, \sovast, 2, 653

\bibitem[{Gradshteyn \& Ryzhik(2007)}]{GradshteynRyzhik2007}
Gradshteyn, I.~S., \& Ryzhik, I.~M. 2007, Table of integrals, series, and products, seventh edn. (Elsevier/Academic Press, Amsterdam), xlviii+1171

\bibitem[{{Graham} \& {Cairns}(2013)}]{GrahamCairns2013}
{Graham}, D.~B., \& {Cairns}, I.~H. 2013, \jgr, 118, 3968, \dodoi{10.1002/jgra.50402}

\bibitem[{{Hinkel-Lipsker} {et~al.}(1989){Hinkel-Lipsker}, {Fried}, \& {Morales}}]{Hinkel-Lipsker1989}
{Hinkel-Lipsker}, D.~E., {Fried}, B.~D., \& {Morales}, G.~J. 1989, \prl, 62, 2680, \dodoi{10.1103/PhysRevLett.62.2680}

\bibitem[{{Hinkel-Lipsker} {et~al.}(1991){Hinkel-Lipsker}, {Fried}, \& {Morales}}]{Hinkel-Lipsker1991}
---. 1991, \prl, 66, 1862, \dodoi{10.1103/PhysRevLett.66.1862}

\bibitem[{{Jebaraj} {et~al.}(2023){Jebaraj}, {Krasnoselskikh}, {Pulupa}, {Magdalenic}, \& {Bale}}]{Jebaraj2023a}
{Jebaraj}, I.~C., {Krasnoselskikh}, V., {Pulupa}, M., {Magdalenic}, J., \& {Bale}, S.~D. 2023, \apjl, 955, L20, \dodoi{10.3847/2041-8213/acf857}

\bibitem[{{Kellogg} {et~al.}(2013){Kellogg}, {Goetz}, {Monson}, \& {Opitz}}]{Kellogg2013}
{Kellogg}, P.~J., {Goetz}, K., {Monson}, S.~J., \& {Opitz}, A. 2013, \jgr, 118, 4766, \dodoi{10.1002/jgra.50443}

\bibitem[{{Kim} {et~al.}(2007){Kim}, {Cairns}, \& {Robinson}}]{Kim2007}
{Kim}, E.-H., {Cairns}, I.~H., \& {Robinson}, P.~A. 2007, \prl, 99, 015003, \dodoi{10.1103/PhysRevLett.99.015003}

\bibitem[{{Kim} {et~al.}(2008){Kim}, {Cairns}, \& {Robinson}}]{Kim2008}
---. 2008, Phys. Plasmas, 15, 102110, \dodoi{10.1063/1.2994719}

\bibitem[{{Krafft} \& {Savoini}(2022)}]{KrafftSavoini2022a}
{Krafft}, C., \& {Savoini}, P. 2022, \apjl, 924, L24, \dodoi{10.3847/2041-8213/ac46a7}

\bibitem[{{Krafft} \& {Savoini}(2024)}]{KrafftSavoini2024}
---. 2024, \apjl, 964, L30, \dodoi{10.3847/2041-8213/ad3449}

\bibitem[{{Krafft} {et~al.}(2024){Krafft}, {Savoini}, \& {Polanco-Rodríguez}}]{Krafft2024}
{Krafft}, C., {Savoini}, P., \& {Polanco-Rodríguez}, F.~J. 2024, \apjl, 967, L20, \dodoi{10.3847/2041-8213/ad47b5}

\bibitem[{{Krafft} \& {Volokitin}(2021)}]{KrafftVolokitin2021}
{Krafft}, C., \& {Volokitin}, A.~S. 2021, \apj, 923, 103, \dodoi{10.3847/1538-4357/ac2153}

\bibitem[{{Krafft} \& {Volokitin}(2024)}]{KrafftVolokitin2024}
---. 2024, \apj, 964, 65, \dodoi{10.3847/1538-4357/ad20ee}

\bibitem[{Krafft {et~al.}(2019)Krafft, Volokitin, \& Gauthier}]{Krafft2019}
Krafft, C., Volokitin, A.~S., \& Gauthier, G. 2019, Fluids, 4, \dodoi{10.3390/fluids4020069}

\bibitem[{{Krafft} {et~al.}(2025){Krafft}, Volokitin, {Polanco-Rodríguez}, \& {Savoini}}]{Krafft2025}
{Krafft}, C., Volokitin, A.~S., {Polanco-Rodríguez}, F.~J., \& {Savoini}, P. 2025, Nature Astronomy, \textit{in press}.
\newblock \url{https://arxiv.org/abs/2506.16816}

\bibitem[{{Krasnoselskikh} {et~al.}(2019){Krasnoselskikh}, {Voshchepynets}, \& {Maksimovic}}]{Krasnoselskikh2019}
{Krasnoselskikh}, V., {Voshchepynets}, A., \& {Maksimovic}, M. 2019, \apj, 879, 51, \dodoi{10.3847/1538-4357/ab22bf}

\bibitem[{{Krasnoselskikh} \& {Sotnikov}(1977)}]{KrasnoselskikhSotnikov1977}
{Krasnoselskikh}, V.~V., \& {Sotnikov}, V.~I. 1977, Fizika Plazmy, 3, 872

\bibitem[{{Krupar} {et~al.}(2015){Krupar}, {Kontar}, {Soucek}, {Santolik}, {Maksimovic}, \& {Kruparova}}]{Krupar2015}
{Krupar}, V., {Kontar}, E.~P., {Soucek}, J., {et~al.} 2015, \aap, 580, A137, \dodoi{10.1051/0004-6361/201425308}

\bibitem[{{Krupar} {et~al.}(2020){Krupar}, {Szabo}, {Maksimovic}, {Kruparova}, {Kontar}, {Balmaceda}, {Bonnin}, {Bale}, {Pulupa}, {Malaspina}, {Bonnell}, {Harvey}, {Goetz}, {Dudok de Wit}, {MacDowall}, {Kasper}, {Case}, {Korreck}, {Larson}, {Livi}, {Stevens}, {Whittlesey}, \& {Hegedus}}]{Krupar2020}
{Krupar}, V., {Szabo}, A., {Maksimovic}, M., {et~al.} 2020, \apjs, 246, 57, \dodoi{10.3847/1538-4365/ab65bd}

\bibitem[{{Krupar} {et~al.}(2024{\natexlab{a}}){Krupar}, {Kruparova}, {Szabo}, {Nemec}, {Maksimovic}, {Martinez Oliveros}, {Lario}, {Bonnin}, {Vecchio}, {Pulupa}, \& {Bale}}]{Krupar2024a}
{Krupar}, V., {Kruparova}, O., {Szabo}, A., {et~al.} 2024{\natexlab{a}}, \apj, 961, 88, \dodoi{10.3847/1538-4357/ad12ba}

\bibitem[{{Krupar} {et~al.}(2024{\natexlab{b}}){Krupar}, {Kruparova}, {Szabo}, {Wilson}, {Nemec}, {Santolik}, {Pulupa}, {Issautier}, {Bale}, \& {Maksimovic}}]{Krupar2024b}
---. 2024{\natexlab{b}}, \apjl, 967, L32, \dodoi{10.3847/2041-8213/ad4be7}

\bibitem[{{Layden} {et~al.}(2013){Layden}, {Cairns}, {Li}, \& {Robinson}}]{Layden2013}
{Layden}, A., {Cairns}, I.~H., {Li}, B., \& {Robinson}, P.~A. 2013, \prl, 110, 185001, \dodoi{10.1103/PhysRevLett.110.185001}

\bibitem[{Lee {et~al.}(2022)Lee, Yoon, Lee, \& Tu}]{Lee2022}
Lee, S.-Y., Yoon, P.~H., Lee, E., \& Tu, W. 2022, The Astrophysical Journal, 924, 36, \dodoi{10.3847/1538-4357/ac32bb}

\bibitem[{Lee {et~al.}(2019)Lee, Ziebell, Yoon, Gaelzer, \& Lee}]{Lee2019}
Lee, S.-Y., Ziebell, L.~F., Yoon, P.~H., Gaelzer, R., \& Lee, E.~S. 2019, \apj, 871, 74, \dodoi{10.3847/1538-4357/aaf476}

\bibitem[{Li {et~al.}(2008{\natexlab{a}})Li, Cairns, \& Robinson}]{Li2008a}
Li, B., Cairns, I.~H., \& Robinson, P.~A. 2008{\natexlab{a}}, \jgr Space Physics, 113, 1, \dodoi{10.1029/2007JA012958}

\bibitem[{Li {et~al.}(2008{\natexlab{b}})Li, Robinson, \& Cairns}]{Li2008b}
Li, B., Robinson, P.~A., \& Cairns, I.~H. 2008{\natexlab{b}}, \jgr Space Physics, 113, 1, \dodoi{10.1029/2008JA013255}

\bibitem[{{Li} {et~al.}(2005){Li}, {Willes}, {Robinson}, \& {Cairns}}]{Li2005}
{Li}, B., {Willes}, A.~J., {Robinson}, P.~A., \& {Cairns}, I.~H. 2005, Phys. Plasmas, 12, 052324, \dodoi{10.1063/1.1906214}

\bibitem[{Lorfing {et~al.}(2023)Lorfing, Reid, G{\'o}mez-Herrero, Maksimovic, Nicolaou, Owen, Rodriguez-Pacheco, Ryan, Trotta, \& Verscharen}]{Lorfing2023}
Lorfing, C.~Y., Reid, H. A.~S., G{\'o}mez-Herrero, R., {et~al.} 2023, \apj, 959, 128, \dodoi{10.3847/1538-4357/ad0be3}

\bibitem[{Malaspina {et~al.}(2012)Malaspina, Cairns, \& Ergun}]{Malaspina2012}
Malaspina, D.~M., Cairns, I.~H., \& Ergun, R.~E. 2012, \apj, 755, 45, \dodoi{10.1088/0004-637X/755/1/45}

\bibitem[{{Melrose}(1980)}]{Melrose1980}
{Melrose}, D.~B. 1980, \ssr, 26, 3, \dodoi{10.1007/BF00212597}

\bibitem[{{Melrose} \& {Sy}(1972)}]{MelroseSy1972}
{Melrose}, D.~B., \& {Sy}, W.~N. 1972, Australian Journal of Physics, 25, 387, \dodoi{10.1071/PH720387}

\bibitem[{{M{\"u}ller} {et~al.}(2020){M{\"u}ller}, {St. Cyr}, {Zouganelis}, {Gilbert}, {Marsden}, {Nieves-Chinchilla}, {Antonucci}, {Auch{\`e}re}, {Berghmans}, {Horbury}, {Howard}, {Krucker}, {Maksimovic}, {Owen}, {Rochus}, {Rodriguez-Pacheco}, {Romoli}, {Solanki}, {Bruno}, {Carlsson}, {Fludra}, {Harra}, {Hassler}, {Livi}, {Louarn}, {Peter}, {Sch{\"u}hle}, {Teriaca}, {del Toro Iniesta}, {Wimmer-Schweingruber}, {Marsch}, {Velli}, {De Groof}, {Walsh}, \& {Williams}}]{Muller2020}
{M{\"u}ller}, D., {St. Cyr}, O.~C., {Zouganelis}, I., {et~al.} 2020, \aap, 642, A1, \dodoi{10.1051/0004-6361/202038467}

\bibitem[{{Muschietti} {et~al.}(1985){Muschietti}, {Goldman}, \& {Newman}}]{Muschietti1985}
{Muschietti}, L., {Goldman}, M.~V., \& {Newman}, D. 1985, \solphys, 96, 181, \dodoi{10.1007/BF00239800}

\bibitem[{{Nishikawa} \& {Ryutov}(1976)}]{NishikawaRyutov1976}
{Nishikawa}, K., \& {Ryutov}, D.~D. 1976, Journal of the Physical Society of Japan, 41, 1757, \dodoi{10.1143/JPSJ.41.1757}

\bibitem[{{Papadopoulos} {et~al.}(1974){Papadopoulos}, {Goldstein}, \& {Smith}}]{Papadopoulos1974}
{Papadopoulos}, K., {Goldstein}, M.~L., \& {Smith}, R.~A. 1974, \apj, 190, 175, \dodoi{10.1086/152862}

\bibitem[{Polanco-Rodríguez {et~al.}(2025)Polanco-Rodríguez, Krafft, \& Savoini}]{Polanco2025a}
Polanco-Rodríguez, F.~J., Krafft, C., \& Savoini, P. 2025, The Astrophysical Journal Letters, 982, L24, \dodoi{10.3847/2041-8213/adba64}

\bibitem[{{Ratcliffe} {et~al.}(2014){Ratcliffe}, {Kontar}, \& {Reid}}]{RatcliffeReid2014}
{Ratcliffe}, H., {Kontar}, E.~P., \& {Reid}, H.~A.~S. 2014, \aap, 572, A111, \dodoi{10.1051/0004-6361/201423731}

\bibitem[{{Reid} \& {Kontar}(2021)}]{ReidKontar2021}
{Reid}, H. A.~S., \& {Kontar}, E.~P. 2021, Nature Astronomy, 5, 796, \dodoi{10.1038/s41550-021-01370-8}

\bibitem[{{Reid} \& {Ratcliffe}(2014)}]{ReidRatcliffe2014}
{Reid}, H. A.~S., \& {Ratcliffe}, H. 2014, Res. Astron. Astrophys., 14, 773, \dodoi{10.1088/1674-4527/14/7/003}

\bibitem[{{Rhee} {et~al.}(2009){Rhee}, {Ryu}, {Woo}, {Kaang}, {Yi}, \& {Yoon}}]{Rhee2009}
{Rhee}, T., {Ryu}, C.-M., {Woo}, M., {et~al.} 2009, \apj, 694, 618, \dodoi{10.1088/0004-637X/694/1/618}

\bibitem[{{Schleyer} {et~al.}(2013){Schleyer}, {Cairns}, \& {Kim}}]{Schleyer2013}
{Schleyer}, F., {Cairns}, I.~H., \& {Kim}, E.~H. 2013, Physics of Plasmas, 20, 032101, \dodoi{10.1063/1.4793726}

\bibitem[{{Schleyer} {et~al.}(2014){Schleyer}, {Cairns}, \& {Kim}}]{Schleyer2014}
{Schleyer}, F., {Cairns}, I.~H., \& {Kim}, E.-H. 2014, Journal of Geophysical Research (Space Physics), 119, 3392, \dodoi{10.1002/2013JA019364}

\bibitem[{{Shafranov}(1967)}]{Shafranov1967}
{Shafranov}, V.~D. 1967, Reviews of Plasma Physics, 3, 1, \dodoi{10.1007/978-1-4615-7799-7}

\bibitem[{{Thejappa} \& {MacDowall}(2021)}]{ThejappaMacDowall2021}
{Thejappa}, G., \& {MacDowall}, R.~J. 2021, \apj, 912, 61, \dodoi{10.3847/1538-4357/abee74}

\bibitem[{{Tsytovich}(1970)}]{Tsytovich1970}
{Tsytovich}, V.~N. 1970, {Nonlinear Effects in Plasma} (Plenum Press, New-York-London)

\bibitem[{{van Haarlem} {et~al.}(2013){van Haarlem}, {Wise}, {Gunst}, {Heald}, {McKean}, {Hessels}, {de Bruyn}, {Nijboer}, {Swinbank}, {Fallows}, {Brentjens}, {Nelles}, {Beck}, {Falcke}, {Fender}, {H{\"o}randel}, {Koopmans}, {Mann}, {Miley}, {R{\"o}ttgering}, {Stappers}, {Wijers}, {Zaroubi}, {van den Akker}, {Alexov}, {Anderson}, {Anderson}, {van Ardenne}, {Arts}, {Asgekar}, {Avruch}, {Batejat}, {B{\"a}hren}, {Bell}, {Bell}, {van Bemmel}, {Bennema}, {Bentum}, {Bernardi}, {Best}, {B{\^\i}rzan}, {Bonafede}, {Boonstra}, {Braun}, {Bregman}, {Breitling}, {van de Brink}, {Broderick}, {Broekema}, {Brouw}, {Br{\"u}ggen}, {Butcher}, {van Cappellen}, {Ciardi}, {Coenen}, {Conway}, {Coolen}, {Corstanje}, {Damstra}, {Davies}, {Deller}, {Dettmar}, {van Diepen}, {Dijkstra}, {Donker}, {Doorduin}, {Dromer}, {Drost}, {van Duin}, {Eisl{\"o}ffel}, {van Enst}, {Ferrari}, {Frieswijk}, {Gankema}, {Garrett}, {de Gasperin}, {Gerbers}, {de Geus}, {Grie{\ss}meier}, {Grit}, {Gruppen}, {Hamaker}, {Hassall}, {Hoeft}, {Holties}, {Horneffer}, {van der Horst}, {van Houwelingen}, {Huijgen}, {Iacobelli}, {Intema}, {Jackson}, {Jelic}, {de Jong}, {Juette}, {Kant}, {Karastergiou}, {Koers}, {Kollen}, {Kondratiev}, {Kooistra}, {Koopman}, {Koster}, {Kuniyoshi}, {Kramer}, {Kuper}, {Lambropoulos}, {Law}, {van Leeuwen}, {Lemaitre}, {Loose}, {Maat}, {Macario}, {Markoff}, {Masters}, {McFadden}, {McKay-Bukowski}, {Meijering}, {Meulman}, {Mevius}, {Middelberg}, {Millenaar}, {Miller-Jones}, {Mohan}, {Mol}, {Morawietz}, {Morganti}, {Mulcahy}, {Mulder}, {Munk}, {Nieuwenhuis}, {van Nieuwpoort}, {Noordam}, {Norden}, {Noutsos}, {Offringa}, {Olofsson}, {Omar}, {Orr{\'u}}, {Overeem}, {Paas}, {Pandey-Pommier}, {Pandey}, {Pizzo}, {Polatidis}, {Rafferty}, {Rawlings}, {Reich}, {de Reijer}, {Reitsma}, {Renting}, {Riemers}, {Rol}, {Romein}, {Roosjen}, {Ruiter}, {Scaife}, {van der Schaaf}, {Scheers}, {Schellart}, {Schoenmakers}, {Schoonderbeek}, {Serylak}, {Shulevski}, {Sluman}, {Smirnov}, {Sobey}, {Spreeuw}, {Steinmetz}, {Sterks}, {Stiepel}, {Stuurwold}, {Tagger}, {Tang}, {Tasse}, {Thomas}, {Thoudam}, {Toribio}, {van der Tol}, {Usov}, {van Veelen}, {van der Veen}, {ter Veen}, {Verbiest}, {Vermeulen}, {Vermaas}, {Vocks}, {Vogt}, {de Vos}, {van der Wal}, {van Weeren}, {Weggemans}, {Weltevrede}, {White}, {Wijnholds}, {Wilhelmsson}, {Wucknitz}, {Yatawatta}, {Zarka}, {Zensus}, \& {van Zwieten}}]{VanHaarlem2013}
{van Haarlem}, M.~P., {Wise}, M.~W., {Gunst}, A.~W., {et~al.} 2013, \aap, 556, A2, \dodoi{10.1051/0004-6361/201220873}

\bibitem[{{Volokitin} \& {Krafft}(2018)}]{VolokitinKrafft2018}
{Volokitin}, A.~S., \& {Krafft}, C. 2018, \apj, 868, 104, \dodoi{10.3847/1538-4357/aae7cc}

\bibitem[{{Volokitin} \& {Krafft}(2020)}]{VolokitinKrafft2020}
---. 2020, \apjl, 893, L47, \dodoi{10.3847/2041-8213/ab74de}

\bibitem[{Willes \& Cairns(2000)}]{WillesCairns2000}
Willes, A.~J., \& Cairns, I.~H. 2000, Physics of Plasmas, 7, 3167, \dodoi{10.1063/1.874180}

\bibitem[{{Yin} {et~al.}(1998){Yin}, {Ashour-Abdalla}, {El-Alaoui}, {Bosqued}, \& {Bougeret}}]{Yin1998}
{Yin}, L., {Ashour-Abdalla}, M., {El-Alaoui}, M., {Bosqued}, J.~M., \& {Bougeret}, J.~L. 1998, \grl, 25, 2609, \dodoi{10.1029/98GL01989}

\bibitem[{Zhou {et~al.}(2020)Zhou, Mu{\~n}oz, B{\"u}chner, \& Liu}]{Zhou2020}
Zhou, X., Mu{\~n}oz, P.~A., B{\"u}chner, J., \& Liu, S. 2020, The Astrophysical Journal, 891, 92, \dodoi{10.3847/1538-4357/ab6a0d}

\bibitem[{{Ziebell} {et~al.}(2015){Ziebell}, {Yoon}, {Petruzzellis}, {Gaelzer}, \& {Pavan}}]{Ziebell2015}
{Ziebell}, L.~F., {Yoon}, P.~H., {Petruzzellis}, L.~T., {Gaelzer}, R., \& {Pavan}, J. 2015, \apj, 806, 237, \dodoi{10.1088/0004-637X/806/2/237}

\bibitem[{{Zlotnik}(1981)}]{Zlotnik1981}
{Zlotnik}, E.~I. 1981, \aap, 101, 250

\end{thebibliography}

\end{document}